\documentclass[12pt]{article}
\usepackage[dvips]{graphicx}

\parskip=\medskipamount

\def\be{\begin{equation}}
\def\ee{\end{equation}}
\def\bea{\begin{eqnarray}}
\def\eea{\end{eqnarray}}
\def\bean{\begin{eqnarray*}}
\def\eean{\end{eqnarray*}}
\def\ve{\varepsilon}

\title{Path integral derivations of\\ novel complex trajectory methods}

\author{Jeremy Schiff${}^{1}$, Yair Goldfarb${}^2$ and David J. Tannor${}^2$\\ 
${}^1${\small{Department of Mathematics, Bar-Ilan University, Ramat Gan, 52900 Israel}}\\
${}^2${\small{Department of Chemical Physics, The Weizmann Institute of Science, Rehovot, 76100 Israel}}} 

\begin{document}

\maketitle

\begin{abstract} 
\noindent Path integral derivations are presented
for two recently developed complex trajectory techniques for the 
propagation of wave packets, Complex WKB and BOMCA. Complex WKB 
is derived using a standard saddle point approximation of the 
path integral, but taking into account the $\hbar$ dependence of 
both the amplitude and the phase of the 
intial wave function, thus giving rise to the need for {\em complex}
classical trajectories. BOMCA is derived using a modification
of the saddle point technique, in which the path integral is approximated 
by expanding around a near-classical path, chosen so that 
up to some predetermined order there is no need to add any  correction 
terms to the leading order approximation. Both Complex WKB and BOMCA
give the same leading order approximation; in Complex WKB higher accuracy is 
achieved by adding correction terms, while in BOMCA no additional terms are ever
added --- higher accuracy is achieved by
changing the path along which the original approximation is computed. 
The path integral derivation of the methods explains the need to 
incorporate contributions from more than one trajectory, as observed in 
previous numerical work. On the other hand, it emerges that the methods provide 
efficient schemes for computing the higher order terms in the asymptotic 
evaluation of path integrals. The understanding we develop of BOMCA suggests
that there should exist near-classical trajectories that give  {\em exact} 
quantum dynamical results  when used in the computation of the path integral keeping
just the leading order term. We also apply our path integral techniques
to give a compact derivation of the semiclassical approximation to the 
coherent state propagator.
\end{abstract}

\section{Introduction} 

In a recent series of papers \cite{bomca1, bomca2, cwkb}
we have considered two complex trajectory techniques
for solving the time-dependent Schr\"odinger equation (TDSE). 
By a ``trajectory technique'' we mean 
that we solve the TDSE for the wave function by integrating a system of ODEs
along certain trajectories in configuration space. By a ``complex trajectory technique'' 
we mean that the relevant trajectories evolve in complex 
configuration space --- 
i.e. we analytically continue the wave function and consider it as a
function of complex space variables. (Note that the time variable remains real,
so the trajectories are real curves in complex space.) The motivation
for using complex trajectories comes from the substitution 
\be \psi=\exp\left(\frac{iS}{\hbar}\right)\ , \quad S\in{\bf C}\ , \label{subs}\ee
in the TDSE
\be i\hbar\psi_t = -\frac{\hbar^2}{2m} \nabla^2 \psi + V({\bf x})\psi\ . \label{TDSE}\ee
This yields the {\em complex} quantum Hamilton Jacobi equation  (CQHJE) \cite{LP,John}
\be
S_t + \frac{1}{2m}\left(\nabla S\right)^2 + V({\bf x}) =
   \frac{i\hbar}{2m} \nabla^2 S\ .
\label{CQHJE} \ee
Taking $\hbar$ as small, the CQHJE can be considered as a perturbation of the 
classical Hamilton Jacobi equation (HJE) 
\be
S_t + \frac{1}{2m}\left(\nabla S\right)^2 + V({\bf x}) = 0\ .
\ee
Since the classical HJE can  be solved exactly by integration along trajectories
in space defined by 
\be \frac{d{\bf x}}{dt} = \frac{\nabla S}{m}\ , \label{traj111}\ee 
it is natural to try a similar technique (at least 
as an approximation) for its perturbation, the CQHJE. 
Complex trajectories arise since $S$ in 
equation (\ref{subs}), and hence $\nabla S$ in equation (\ref{traj111}), are
complex, leading to 
complex initial conditions for the evolution; furthermore
the perturbation in (\ref{CQHJE}) is complex.

In our earlier papers we observed that a reasonable 
approximation to the wave function 
may require taking into account
contributions from  more than one trajectory  reaching a 
particular point in space. We gave no theoretical justification for this, 
and one of the purposes of the current paper is to fill this hole.
More generally the aim of the current paper is to strengthen 
the theoretical basis of the techniques of our previous papers by
showing how they can be derived by saddle point
evaluation of the wave function in the path 
integral representation. In this approach the 
need to (potentially) add the contributions of several trajectories 
emerges naturally. 
 
In section 2 we give a detailed presentation of the two techniques 
under study here. In the first method,
{\sl Complex WKB} \cite{cwkb}, the trajectories are solutions of the 
classical equations of motion. In the second method,
which we call {\sl BOMCA} \cite{bomca1,bomca2},
(BOhmian Mechanics with Complex Action), the trajectories are 
order $\hbar$ perturbations of solutions of the classical equations. 
Indeed, in BOMCA the  trajectories depend on 
the order in $\hbar$, while in complex WKB 
they remain the same; to increase the order in $\hbar$ in complex WKB we have 
to integrate a further system of ODEs along the trajectories. 

In section 3 we present a path integral derivation of Complex WKB. 
Since the trajectories involved in Complex WKB are classical, 
this involves a standard saddle point approximation of the path integral.
The approach, however, is still nonstandard in that the initial value 
of the wave function is taken into account, leading to
complex classical trajectories.  In this regard, our approach 
differs from standard time-dependent WKB theory \cite{tdwkb}. Our approach is 
the appropriate one when the initial wave function has the form $\exp
(iS^{\rm init}({\bf x})/\hbar)$, as implied by (\ref{subs})
(though note an interesting recent paper of Maia {\it et al.} \cite{Maia}). 
The equivalence of  Complex WKB to the saddle point approximation of the path integral 
gives  rise to a potentially very useful result.
While it has long been recognized that certain factors 
involved in the (leading order) saddle point approximation of path integrals, 
specifically elements of the so-called stability matrix,
can be calculated  efficiently by integrating certain ODEs along classical 
trajectories \cite{smes}, it turns out that the same is true (at least in our 
situation) for all higher order 
correction terms.  Within the path integral formulation, the 
expressions for higher order correction terms 
involve complicated multiple integrals; the Complex WKB method
reformulates these expressions as solutions of a system of first order ODEs, 
which is much easier to handle computationally.

Before presenting the path integral derivation of BOMCA, in section 4
we present a slight modification to the standard method of asymptotic
analysis of integrals with a large parameter. In section 5 we apply this
modification to the path integral, and are led to BOMCA. The distinction 
between BOMCA and Complex WKB becomes extremely clear. Complex WKB
and BOMCA  give the  same leading order approximation to the wave function 
$\psi({\bf X},T)$, 
determined as follows: First find complex classical trajectories ${\bf x}(t)$ 
satisfying appropriate initial and final conditions, specifically, solve the problem
\be
m\ddot{\bf x}(t)+\nabla V({\bf x}(t))=0 \ , 
\qquad  
\dot{\bf x}(0)=-\frac{i\hbar}{m} \nabla \log\psi_0({\bf x}(0)) \ , 
\quad
{\bf x}(T)={\bf X}\ . 
\label{class}\ee
Here $\psi_0$ is the wave function  at $t=0$. 
Next, for each such trajectory, compute the matrix $U$ satisfying 
\be 
m \ddot{U} + H(V)({\bf x}(t)) U  = 0 \ , \qquad 
U(0)=I\ , \qquad \dot{U}(0)=-\frac{i\hbar}{m}H(\log\psi_0)({\bf x}(0)) \ .
\label{jac}\ee 
Here $H(V)$ denotes the matrix of second derivatives of $V$ and $H(\log\psi_0)$
the matrix of second derivatives of $\log\psi_0$. 
Then the wave function is approximated by 
\be \psi({\bf X},T) \approx \sum 
    \frac{e^{iS[{\bf x}]/\hbar}\psi_0({\bf x}(0))}{\sqrt{\det U(T)}} \ ,
\label{clwf}\ee
where $S[{\bf x}]$ denotes the classical action associated with the path ${\bf x}(t)$,
i.e.
\be 
S[{\bf x}] = \int_0^T \left(  \frac12 m \dot{\bf x}^2 - V({\bf x}) \right) \ dt\ .  
\label{act}\ee   
The sum in (\ref{clwf}) is over contributing trajectories 
(possibly not all trajectories), as we will explain later. 
The distinction between Complex WKB and BOMCA lies in the manner in which 
higher order corrections are made to (\ref{clwf}). In Complex WKB higher order corrections 
are made  by multiplying the leading order contribution for each trajectory 
in (\ref{clwf}) by a suitable factor of the form $1+O(\hbar)$. 
In BOMCA, the formula (\ref{clwf}) is never modified, but the paths 
${\bf x}(t)$ and matrices $U(t)$ are no longer required to be classical.
More explicitly, the differential equations in (\ref{class}) and (\ref{jac}) are
replaced by equations of the form
\bea
m\ddot{\bf x} + \nabla V({\bf x}(t)) &=& O(\hbar)\ , \label{pertx}\\  
m\ddot{U} + H(V)({\bf x}(t)) U  &=& O(\hbar)\ .  \label{pertU}
\eea 
BOMCA gives explicit expressions for the terms to introduce on the right hand side 
of these equations, but, as we shall explain, they are not unique choices. 

We call the quantity appearing on the right hand side of (\ref{clwf}), with the 
classical choice of ${\bf x}$ and $U$, the {\sl classical wave function}. 
Note that our use of the term ``classical wave function'' differs from
previous uses, see for example Box 2.2 in \cite{Wyatt}. We emphasize also 
that our classical wave function differs from the usual approximations made
in time dependent WKB theory;
the difference can be traced to different assumptions about the $\hbar$ dependence of the 
initial wave function, with our choice requiring the use of  complex trajectories. 

As we have explained, BOMCA provides a prescription for making  
the formula (\ref{clwf}) more accurate, to any order in $\hbar$, by changing the equations 
that  ${\bf x}$ and $U$ satisfy. We are led to conjecture that 
there may exist choices of ${\bf x}$ and $U$, satisfying 
(\ref{pertx})-(\ref{pertU}), such that formula (\ref{clwf}) is {\em exact}. Unfortunately, 
at this stage we only know how to describe the right hand sides of   
equations (\ref{pertx}) and (\ref{pertU}) perturbatively in $\hbar$, and, as 
we have indicated above, there are many choices (one being associated with 
BOMCA). If there exist choices of ${\bf x}$ and 
$U$ for which (\ref{clwf}) is exact, the relevant trajectories ${\bf x}$ would be an 
interesting  intermediate object between classical and quantum trajectories. 
The usual notion
of quantum trajectories (in Bohmian mechanics) are the paths in (real) configuration 
space satisfying $\dot{\bf x}=\frac{\hbar}{m} {\rm Im} \left( \nabla \log\psi({\bf x},t)
\right)$ (see the books \cite{Holland} and \cite{Wyatt} for extensive   
discussion). One of the properties of these trajectories is that the velocity 
diverges at a node of the wave function, so near nodes 
quantum trajectories are qualitatively different from classical 
trajectories. In distinction to this, 
the non-classical trajectories that arise in BOMCA  are always
perturbations of classical trajectories. Certainly it is 
possible to express the wave function $\psi$ in the form 
(\ref{clwf}) only in certain regions of 
${\bf X},T$ space (like any other semiclassical formula, our
formula suffers from problems related to caustics and Stokes' lines), but in these 
regions we conjecture that there exist non-classical trajectories which are 
perturbations of classical trajectories, which make the formula (\ref{clwf}) exact. 

After our derivation of BOMCA from the path integral, in section 6 we 
discuss the application of our ideas to the evaluation of other quantities in quantum
mechanics. There is an extensive literature on the use of complex 
classical trajectories to compute the coherent state propagator, 
(see for example \cite{Klauder1,Klauder2,Weissman,Adachi,XdA1,XdA2,XdA3,VVH1,VVH2,deA1,deA2}),  
and  we show how some relevant formulae can be derived using our techniques. 
Section 7 contains concluding comments. An appendix
provides  the multidimensional derivation of the classical wave function (\ref{clwf});
in the most of the main text we present the path integral derivations just 
in the one dimensional case. 

We conclude this introduction with a discussion of some relevant literature that
has not yet been mentioned.  The use of complex classical trajectories 
in semiclassical quantum mechanics goes back to Stine and Marcus \cite{Stine}
and  Miller and George \cite{Miller1,Miller2,Miller3}, and numerous different applications have 
been subsequently presented (see for example \cite{Shudo,Kus}).  As far as we know, the 
first attempt
to use complex classical trajectories to propagate wave packets  
is in the works of Huber, Heller and Littlejohn \cite{HH,HHL} (the 
superposition of 
contributions from  more than one trajectory also appears in this work).  
Our work, however, is closer to
the rather different viewpoint of Boiron and Lombardi \cite{BL}. Other developments in this
area include the extensive work of de Aguiar and  collaborators \cite{deA22,deA23,deA24}, 
and very recent contributions of Chou and Wyatt \cite{CW} and 
Sanz and Miret-Artes \cite{SM}. There are a number of   papers 
on time dependent WKB that we also found illuminating: \cite{ORW,Bracken,Gorman,SB}. 
Recent numerical work of Bender, Brody and Hook \cite{Bender}, 
suggesting strong connections between 
complex classical dynamics and quantum dynamics, is very encouraging; such connections 
also provided the motivation for the detailed studies of complex classical dynamics
of Kay and Shnerb \cite{Kay1,Kay2}. Likewise, there are interesting connections 
between our current work
and a series of papers by Poirier and collaborators
\cite{B1,B2,B3,B4,B5,B6}. Poirier  considers 
a decomposition of the wave function as a sum of (nodeless) terms;  
we suspect this decomposition 
is strongly linked, if not identical to, the decomposition implied
by (\ref{clwf}) (at least in the time-dependent case \cite{B5}).
Finally, we note that the hierarchy of ODEs in the 
BOMCA method can be viewed as a complex version of 
the DPM of Trahan, Hughes and Wyatt \cite{THW} 
which was originally developed for real trajectories (see the exposition in \cite{Wyatt}
and the comparison of BOMCA and DPM in \cite{yair}). 


\section{The Complex WKB and BOMCA methods}

As explained in the introduction, the starting point for both 
the methods we consider in this paper is the substitution 
$\psi=e^{iS/\hbar}$ in the time-dependent Schr\"odinger equation
to obtain the CQHJE (\ref{CQHJE}). Here $S({\bf x},t)$ is complex.
Both of our methods 
consist of approximating the CQHJE by a system of equations that 
can be solved by integrating along trajectories in complex
configuration space. Both of our methods allow us to systematically
improve the order of approximation in such a way that we might 
reasonably expect, in a suitable limit, to obtain exact results.   

The first method, 
{\em complex WKB}, proceeds by an expansion 
of $S$ in powers of $\hbar$. The relevant trajectories, irrespective 
of the order of approximation, are solutions of the complex
classical equations of motion. Complex WKB is described in detail 
in subsection  2.1.  The second method, {\em BOMCA}, 
involves a different 
approximation scheme that will be described in detail in section 2.2. 
The relevant trajectories depend on the order of approximation. The
question arises as to whether these trajectories have a 
well-defined limit as the order of approximation is increased, and what 
this limit is. We will discuss this matter more in section 5.

In subsection 2.3 we summarize some of the findings of our previous 
work on complex WKB and BOMCA that are relevant for understanding the rest
of this paper. 

\subsection{Complex WKB}

Writing 
\be  S({\bf x},t) = \sum_{n=0}^\infty S_n({\bf x},t) \hbar^n \ee
and substituting in the CQHJE we obtain the following PDEs for 
the component functions $S_n({\bf x},t)$:
\be
S_{0t} + \frac{1}{2m}\left(\nabla S_0\right)^2 + V({\bf x}) = 0 \ , \label{cw0}
\ee
\be
S_{nt} +  \frac{\nabla S_0 }{m} \cdot \nabla S_n 
= 
- \frac{1}{2m}\sum_{m=1}^{n-1} \nabla S_m \cdot \nabla S_{n-m} 
+ \frac{i}{2m} \nabla^2 S_{n-1}\ , \quad n=1,2,\ldots\ . \label{cwn}
\ee
Along with the TDSE we assume we are provided with an initial 
wave function $\psi_0({\bf x})=\psi({\bf x},0)$, and this provides 
us with an initial condition 
$S({\bf x},0)=-i\hbar\log\psi_0({\bf x})$
for the CQHJE. How are we to choose appropriate initial conditions
for the component functions $S_n$? The archetypal form of the initial
wave function we wish to consider  is
a nonnormalized Gaussian wave-packet, which in one spatial dimension  takes  the form 
\be 
\psi_0(x) =  
\exp\left(
-\frac{(a_0+ia_1)(x-x_0)^2}{\hbar} + \frac{ip_0(x-x_0)}{\hbar}
\right)\  .
\label{gwp}\ee
Here $a_0,a_1,x_0,x_1$ are real 
constants, with $a_0>0$, related to the expectations
and variance of  position and momentum via
\bea  
 \langle x\rangle=x_0 \ , &\qquad&  \langle p\rangle=p_0  \ , \\
 \langle (x-x_0)^2\rangle=\frac{\hbar}{2a_0} \ , &\qquad& 
 \langle (p-p_0)^2\rangle=\frac{\hbar(a_0^2+a_1^2)}{2a_0} \ .  
\eea
For this choice of $\psi_0$ we have 
\be 
S(x,0)
=-i\hbar\log\psi_0(x)
=i(a_0+ia_1)(x-x_0)^2 + p_0(x-x_0) \ , 
\ee 
so it is natural to choose 
\bea
S_0(x,0) &=& i(a_0+ia_1)(x-x_0)^2 + p_0(x-x_0) \ , \\
S_n(x,0) &=&  0\ , \qquad n=1,2,\ldots . 
\eea
In greater generality, we assume that 
we can write the initial wave function as 
$\psi_0({\bf x})=\exp(iS^{\rm init}({\bf x})/\hbar)$ 
and take $S_0({\bf x},0)=S^{\rm init}({\bf x})$ and 
$S_n({\bf x},0)=0$ for $n\ge 1$. It is possible to generalize
to the case that $S({\bf x},0)$ can be expanded in a Taylor series
in $\hbar$. 

In complex WKB we opt to 
solve the system of equations (\ref{cw0})-(\ref{cwn}) by integrating
along trajectories defined by 
\be 
\frac{d{\bf x}}{dt} = \frac{\nabla S_0}{m}\ .
\label{xeq} \ee
Writing ${\bf v}=\frac{\nabla S_0}{m}$ and taking the gradient of (\ref{cw0}) gives 
\be
\frac{\partial v_i}{\partial t} + \left({\bf v}\cdot\nabla\right) v_i + 
  \frac{1}{m}\frac{\partial V}{\partial x_i} = 0  \ .
\ee
Thus along the trajectories we have 
\be 
\frac{d{\bf v}}{dt} =  -\frac1{m} \nabla V({\bf x}) \ . 
\ee
(Here $\frac{d}{dt}$ denotes the Lagrangian derivative along the trajectories
$\frac{d}{dt} = \frac{\partial}{\partial t} + {\bf v}\cdot\nabla $.) We see the 
trajectories are simply classical trajectories. Note however that they are 
trajectories in {\em complexified} space. The initial condition for the TDSE
gives the initial condition 
$S^{\rm init}({\bf x})$ as a complex function of ${\bf x}$, and thus
\be {\bf v}(0) = \frac1{m} \nabla S^{\rm init}({\bf x}(0)) 
\label{veq}\ee
is in general complex. 

To summarize to this stage: in complex WKB we choose to integrate along 
trajectories given by (\ref{xeq}), and from (\ref{cw0}) we deduce that these are actually
classical trajectories, i.e. solutions of 
\be 
\frac{d{\bf x}}{dt} = {\bf v}\ , 
\qquad
\frac{d{\bf v}}{dt} =  -\frac1{m} \nabla V({\bf x}) \ ,
\label{Nes}\ee 
with the complex initial condition (\ref{veq}). 
Also from 
(\ref{cw0}) we deduce that the evolution of $S_0$ down these trajectories is
given by 
\be \frac{dS_0}{dt} = \frac12 m{\bf v}^2 - V({\bf x})\ .  \label{S0e}\ee
To compute the evolution of $S_1,S_2,\ldots$ along the trajectories we need 
to use the equations (\ref{cwn}). We have written these equations with precisely
the Lagrangian derivative of $S_n$ along the trajectories on the left hand side. 
Writing the first few equations out more explicitly we have
\bea
\frac{dS_1}{dt} &=& \frac{i}{2m} \nabla^2 S_{0}\ , \label{S1e}\\
\frac{dS_2}{dt} &=& - \frac{1}{2m} \left( \nabla S_1 \right)^2   
                 + \frac{i}{2m} \nabla^2 S_{1} \ , \label{S2e}\\
\frac{dS_3}{dt} &=& - \frac{1}{m} \nabla S_1 \cdot \nabla S_2 
              + \frac{i}{2m} \nabla^2 S_{2}\ .
\eea
We see that to find $S_1$ we need to follow the evolution of 
$\nabla^2 S_{0}$ along the trajectory, which may be found by 
calculating the second spatial derivatives of (\ref{cw0}). 
To find $S_2$, we see from (\ref{S2e}) that we  need to follow the evolution of 
first and second spatial derivatives of $S_1$ along the trajectories. These are
obtained by taking two spatial derivatives of (\ref{S1e}), but to integrate the 
resulting equations we also need third and fourth derivatives of $S_0$, 
obtained by further differentiation of (\ref{cw0}). 

Proceeding in this manner we see that 
to obtain $S_0,S_1,\ldots,S_n$ we need to follow up to the $2n$'th
derivatives of $S_0$, up to the $2(n-1)$'th derivatives of $S_1$ etc. along the 
trajectory. The total number of derivatives of order up to $i$ of a scalar function 
is  
\be
1 + d + \frac{d(d+1)}{2} + \frac{d(d+1)(d+2)}{6} + 
\ldots + \frac{d(d+1)\ldots(d+i-1)}{i!}  
= \frac{(d+i)!}{d!i!} \ ,
\ee
where $d$ denotes the number of spatial dimensions. 
Thus the total number of functions we need to follow along the trajectories
is given by
\be
D(n,d) 
=  \sum_{\stackrel{i=0}{i~{\rm even}}}^{2n} \frac{(d+i)!}{d!i!} 
\ .  
\label{Dnd}\ee
We are not aware of a closed form expression for this sum, but tabulate it
in Table 1 for some low values of $d$ and $n$. For fixed $n$ and large $d$ we
note that $D(n,d)\sim d^{2n}/(2n)!$, i.e. the number of functions we need to
follow increases polynomially with the dimension. 

\begin{table}
\begin{tabular}{c|c|c|c|c}
$D(n,d)$ & $n=1$ & $n=2$ & $n=3$ & general $n$ \\
\hline
$d=1$  &  4   & 9    & 16  & $(n+1)^2$ \\
\hline
$d=2$   & 7   & 22  &  50  & $\frac16(n+1)(n+2)(4n+3)$ \\
\hline
$d=3$   &  11   & 46  & 130 & $\frac16(n+1)(n+2)(2n^2+6n+3)$ \\
\hline 
general $d$ &  $\frac12(d^2+3d+4)$ &  
   $\frac1{24}(d^4+\ldots)$ & $\frac1{720}(d^6+\ldots)$   &  see (\ref{Dnd})
\end{tabular}
\caption{Total number of functions to be evolved along the trajectories as 
a function of dimensionality $d$ and order $n$. The full expressions in the 
cases $n=2$ and $n=3$ with general $d$ are  
$\frac1{24}(d^4+10d^3+47d^2+86d+72)$  and 
$\frac1{720}(d^6+21d^5+205d^4+1035d^3+3034d^2+4344d+2880)$ 
respectively. Note the numbers listed include a contribution of $d$ for 
finding $\nabla S_0=m{\bf v}$, which in practice is already determined 
when finding the trajectories.} 
\end{table}

At this juncture we write out in full 
the equations that we must integrate to
obtain $S_0,S_1,S_2$: To find the trajectories we solve Newton's equations 
(\ref{Nes}) with the required initial condition (\ref{veq}).
The gradient of $S_0$ along the trajectories can be identified with 
$m{\bf v}$, and we do not need to recompute it. Higher derivatives of $S_0$ 
along the trajectories are determined by integrating the equations:
\bea 
\frac{dS_{0,ik}}{dt}    &=& - V_{ik} -  \frac1{m} \sum_j S_{0,ij} S_{0,jk}\ ,   
    \label{matrixre}\\
\frac{dS_{0,ikl}}{dt}   &=& - V_{ikl} 
   -  \frac1{m} \sum_j \left( S_{0,ij} S_{0,jkl} + S_{0,kj} S_{0,jli} +
                              S_{0,lj} S_{0,jik}  \right)\ , \label{mre2}\\
\frac{dS_{0,iklm}}{dt}   &=& - V_{iklm} 
   -  \frac1{m} \sum_j \left( S_{0,ij} S_{0,jklm} + S_{0,kj} S_{0,jlmi} +
                              S_{0,lj} S_{0,jmik} + S_{0,mj} S_{0,jikl}
         \right)\nonumber \\
&&   - \frac1{m} \sum_j \left( 
    S_{0,jkl} S_{0,jim} + S_{0,jkm} S_{0,jil} + S_{0,jlm} S_{0,jik} \right)\ .
 \label{mre3}\eea
Here $V_{ik}$ denotes $\frac{\partial^2V}{\partial x_i\partial x_k}$ etc, 
and $S_{0,ik}$ denotes $\frac{\partial^2S_0}{\partial x_i\partial x_k}$ etc. 
Derivatives of $S_1$ are determined by integrating the equations
\bea
\frac{dS_{1,i}}{dt}  &=&  \frac{i}{2m}\sum_j  S_{0,jji} 
   - \frac1{m} \sum_j S_{0,ij} S_{1,j}\ ,  \label{S11}\\
\frac{dS_{1,ik}}{dt}  &=&  \frac{i}{2m}\sum_j  S_{0,jjik} 
   - \frac1{m} \sum_j S_{0,ikj} S_{1,j} 
-\frac1{m} \sum_j \left( S_{0,kj} S_{1,ij} + S_{0,ij} S_{1,kj} \right)\ . \label{S12}
\eea 
Finally $S_0,S_1,S_2$ are obtained by integrating the equations 
(\ref{S0e}),(\ref{S1e}) and (\ref{S2e}) respectively. The initial conditions for 
all these equations are obtained from the initial condition for the TDSE
via the function $S^{\rm init}({\bf x})$. Explicitly, we have 
\be
\begin{array}{cccc}
S_0(0) = S^{\rm init}({\bf x}(0)) &
S_{0,i}(0) = S_{i}^{\rm init}({\bf x}(0)) &
S_{0,ij}(0) = S_{ij}^{\rm init}({\bf x}(0)) &
\ldots \\
S_1(0) = 0 &
S_{1,i}(0) = 0 &
S_{1,ij}(0) = 0 &
\ldots  \\
S_2(0) = 0 &
S_{2,i}(0) = 0 &
S_{2,ij}(0) = 0 &
\ldots  \\
\vdots & \vdots & \vdots & 
\end{array} 
\ee
Note that for an initial Gaussian wave packet (of the form (\ref{gwp}) in one 
dimension) the function
$S^{\rm init}$ is quadratic in the spatial variable, so only its first 
two derivatives will be nonzero. 

A few  final notes before leaving 
our description of the complex WKB method: First, observe that
when computing to order
$n$ in complex WKB we use the derivatives of the potential up to order $2n$.
We will see later how this emerges from the path integral approach. 
Second, observe that equation (\ref{matrixre}) is an example of a matrix Riccati equation 
\cite{mre}, and in particular, it has solutions that become infinite in finite time. 
These singularities are a manifestation of the phenomenon of {\em caustics}, which 
appear in almost every application of the WKB method. We note however that the 
singularities in the matrix Riccati equation are pole-type singularities, and it 
is possible, in a suitable sense, to integrate through them \cite{ss}. This is reminiscent
of the fact that it is often possible to ``regularize'' caustics 
\cite{regcau1,regcau2,regcau3,regcau4}. We are
currently investigating the singularity structure of the full system of equations 
(\ref{S1e})-(\ref{S2e}),(\ref{matrixre})-(\ref{mre3}),(\ref{S11})-(\ref{S12})
\cite{jsprep}. Finally, we mention
that although in this paper we work with the multidimensional Schr\"odinger equation 
in the form (\ref{TDSE}), assuming the mass matrix to be a multiple of the identity,
there is no problem extending our formalism to work with a general positive definite
mass matrix.

\subsection{BOMCA}  

BOMCA is an alternate trajectory based approach for solving the CQHJE
(\ref{CQHJE}). Unlike complex WKB it does not involve an expansion in powers
of $\hbar$. Another distinction is that in complex WKB the trajectories are
classical paths, and in BOMCA they are not. Furthermore, the trajectories
in BOMCA depend on the order of the approximation. 

In BOMCA we aim to integrate the CQHJE (\ref{CQHJE}) by integrating  along 
trajectories defined by 
\be 
\frac{d{\bf x}}{dt} = {\bf v}\ , \quad {\rm where} \quad
{\bf v}=\frac{\nabla S}{m}\ .
\ee
Differentiating the CQHJE we see that along these trajectories the 
velocity field ${\bf v}$ satisfies
\be 
\frac{d{\bf v}}{dt} = -\frac1{m} \nabla V({\bf x}) + \frac{i\hbar}{2m^2} 
   \nabla^2 \left( \nabla S \right) \ . 
\label{evolv}\ee
From the CQHJE we see that along such trajectories 
\be
\frac{dS}{dt} = \frac1{2m}{\bf v}^2 -V({\bf x}) -\frac{i\hbar}{2m} \nabla^2 S \ .
\label{Sev}\ee
The problem integrating (\ref{evolv}) and (\ref{Sev}) is that we have no information about
the second and third derivatives of $S$ that appear on the right hand sides. Borrowing
an idea from complex WKB, we differentiate the CQHJE to find 
equations for the evolution of second and higher derivatives of $S$
along the trajectories. At this stage we just write the equations for
evolution of second, third and fourth derivatives:
\bea
\frac{dS_{ij}}{dt} &=& -V_{ij} -\frac1{m} \sum_{p} S_{ip}S_{pj} 
                           +  \frac{i\hbar}{2m} \sum_p S_{ijpp}\ , \label{bo2}\\  
\frac{dS_{ijk}}{dt} &=& -V_{ijk} -\frac1{m} \sum_{p} ( S_{ip}S_{pjk} + S_{jp}S_{pki} + S_{kp}S_{pij} )
            + \frac{i\hbar}{2m} \sum_p S_{ijkpp} \ ,\label{bo3}\\
\frac{dS_{ijkl}}{dt} &=& -V_{ijkl} 
         -\frac1{m} \sum_{p} ( S_{ip}S_{pjkl} + S_{jp}S_{pkli} + S_{kp}S_{plij} + S_{lp}S_{pijk} ) 
    \nonumber \\
  &&        -\frac1{m} \sum_{p} ( S_{ijp}S_{pkl} + S_{ikp}S_{pjl} + S_{ilp}S_{pjk} )
+ \frac{i\hbar}{2m} \sum_p S_{ijklpp}\ . \label{bo4}
\eea
Apparently things have not improved: on the right hand sides of these 
equations fifth and sixth derivatives of $S$ appear. Now we can state the 
procedure of the BOMCA approximation: {\em in $n$th order BOMCA  ignore 
all terms involving derivatives of  $S$ of order exceeding $2n$}. Thus, in 
1st order BOMCA the nonclassical term in (\ref{evolv}) is taken to be zero and the 
trajectories are simply classical trajectories. The evolution (\ref{Sev}) for $S$, 
however,  involves a nonclassical term with second derivatives; but these 
second derivatives are computed by integrating (\ref{bo2}) down the trajectories,
after ignoring the term with 4th order derivatives in (\ref{bo2}). A comparison
with the equations of complex WKB establishes that {\em lowest order BOMCA is equivalent
to lowest order complex WKB} (i.e. complex WKB where only the terms $S_0$ and $S_1$ 
are retained.)

Moving on to 2nd order BOMCA, a nonclassical term with third derivatives of $S$ 
now remains in the equation for the trajectories (\ref{evolv}), and fourth derivatives
of $S$ appear in (\ref{bo2}). The evolution of 
the third and fourth derivatives of $S$ is given by (\ref{bo3})-(\ref{bo4}) after
ignorning the higher derivative terms. We 
observe that the resulting equations 
are precisely the same as equations (\ref{mre2})-(\ref{mre3}) that appeared
in complex WKB. The trajectories, however, are different --- thus 2nd order
BOMCA is not equivalent to complex WKB of {\em any} order. The same is true for
higher order BOMCA. Complex WKB and BOMCA however share the property that 
order $n$ calculations  involves derivatives of the potential $V$ of order up to $2n$.

Note that ignoring the 5th and 6th derivative terms in (\ref{bo3})-(\ref{bo4}) gives
rise to order $\hbar$ errors in the 3rd and 4th derivatives of $S$. Through 
equations (\ref{evolv}) and (\ref{bo2}) this gives rise to order $\hbar^2$ errors in 
the trajectory ${\bf x}$ and the second derivative of $S$. At first glance 
it seems that the order $\hbar^2$ error in ${\bf x}$ should give rise to an 
order $\hbar^2$ error in $S$, as calculated from (\ref{Sev}). But a careful calculation 
shows that the errors induced in $S$ by both the error in ${\bf x}$ and the error
in $S_{ij}$ are of order $\hbar^3$, and thus we have achieved second order accuracy in
$S$ (and first order accuracy in $S/\hbar$, which is what determines the wave function). 
A similar calculation shows that in $n$th order BOMCA, as described above, we achieve
$n$th order accuracy in $S$. There is no evident benefit to truncating the BOMCA 
equations, say, by ignoring 4th derivatives  but not 3rd. This point was not 
adequately appreciated in \cite{bomca1,bomca2}. 

For clarity, we collect here the evolution equations for 2nd order BOMCA in the 
one dimensional case:
\bea
S_t &=&  \frac12 m v^2 - V(x) + \frac{i\hbar}{2m}S''\ ,  \label{zzz1}\\
\frac{dx}{dt} &=& v \ ,  \\ 
\frac{dv}{dt} &=&  -\frac1{m} V'(x) + \frac{i\hbar}{2m^2}S'''\ ,    \\ 
\frac{dS''}{dt} &=& -V''(x)  -\frac1{m} (S'')^2   + \frac{i\hbar}{2m}S''''\ ,    \\ 
\frac{dS'''}{dt} &=& -V'''(x)       -\frac3{m} S''S'''\ , \\
\frac{dS''''}{dt} &=& -V''''(x)       -\frac4{m} S''S'''' - \frac{3}{m}(S''')^2  \ ,
\eea
This system is a singular perturbation of the Newton's equations. 
The system can be somewhat simplified. Introducing 
a new variable $f(t)$ defined (up to multiplication by a constant) by 
$S''=\frac{m}{f}\frac{df}{dt}$, we can solve the $S'''$ and $S''''$ evolution
equations and find that the trajectories are determined by 
\bea
m \frac{d^2x}{dt^2} &=& - V'(x(t))  + \frac{i\hbar}{2m} S'''(t)\ ,   \label{zzz2}\\ 
m \frac{d^2f}{dt^2} &=& - V''(x(t))f(t)  + \frac{i\hbar}{2mf(t)^3} \left( L - 
    \int_0^t f(u)^4 \left( V''''(x(u))   +  \frac{3}{m}S'''(u)^2 \right) du 
            \right) \ ,    \label{zzz3}\\
S'''(t)  &=& \frac1{f(t)^3} \left(  K - \int_0^t V'''(x(u)) f(u)^3 du  \right) \ .
\label{zzz4}\eea 
Here $K,L$ are  constants of integration related to $S'''(0),S''''(0)$. 

We have not yet dealt with the question of initial conditions in BOMCA,
but this is straightforward. Writing, as before, 
\be S^{\rm init}({\bf x}) = -i\hbar\log \psi({\bf x},0)\ , \ee
we take 
\bea
S(0)&=&S^{\rm init}({\bf x}(0))\ , \label{ic1}\\
v_i(0)&=& \frac1{m}S^{\rm init}_i({\bf x}(0))\ , \label{ic2}\\  
S_{ij}(0)&=& S^{\rm init}_{ij}({\bf x}(0))  \qquad {\rm etc.} \label{ic3}
\eea
The number of equations is  easier to discuss in BOMCA 
than in complex WKB. In $n$th order BOMCA we retain derivatives of $S$
up to order $2n$, i.e. a total of $\pmatrix{d+2n \cr d\cr}$ functions. We also  
need to integrate to find {\bf x}; thus there are a total of
\be
\pmatrix{d+2n \cr d\cr} + d 
\ee
functions. But this number cannot be directly compared with the number
in complex WKB. In complex WKB first the trajectories are computed 
using Newton's equations.
If the aim is to determine the wave function at {\bf X} at time $T$ we 
solve (\ref{Nes}), with boundary conditions (\ref{veq}) and ${\bf x}(T)={\bf X}$.
Once the trajectories are determined, the evolution of all the other 
functions along the trajectories is computed. In BOMCA it is necessary 
to solve for all the functions (admittedly a rather smaller number) in order to 
determine the trajectories; that is we look for a solution of the full 
BOMCA system satisfying initial conditions (\ref{ic1})-(\ref{ic3}) and the final 
condition ${\bf x}(T)={\bf X}$. At this stage we have not made a complete 
study of the 
relative efficiencies of the two approaches. 

\subsection{The need for multiple trajectories} 

We refer to our previous papers \cite{bomca1},\cite{bomca2},\cite{cwkb}
for full details of the implementation of the methods and explicit  
numerical examples. 
In order to determine the wave function at position  
${\bf X}$ and time $T$, it is necessary to find trajectories ${\bf x}(t)$ satisfying 
all the necessary initial conditions and the condition ${\bf x}(T)={\bf X}$. The missing 
initial data is simply the starting point of the trajectories, ${\bf x}(0)$. In 
every case we investigated we found there were multiple trajectories 
satisfying all the necessary conditions. That is, 
there are various possible choices of ${\bf x}(0)$, and we refer to the 
different possible choices as different ``branches''. 
In certain cases, the wave 
function associated with one branch gives an accurate result. In other 
cases, it is necessary to add the wave functions associated with more than 
one branch to get an accurate result. In still other cases, one branch 
gives an overwhelmingly large contribution and has to be discarded. For 
certain values of ${\bf X}$ and $T$ there are transitions between the different 
behaviors, and in the neighborhoods of such transitions we could not 
get reasonable accuracy with our methods. 

The upshot of all this is that our derivation of the complex WKB and BOMCA
equations starting from the CQHJE (\ref{CQHJE}) is apparently not telling us 
the whole story. In the following sections we will present derivations of 
complex WKB and BOMCA starting from the path integral formulation of 
quantum mechanics. In this approach the existence of multiple branches, and 
the need to sometimes incorporate one, sometimes more, is easily explained. 
We also find a non-technical explanation of what the trajectories in BOMCA
are. At the same time, we see that the equations we have presented in 
detail for complex WKB and BOMCA actually provide an efficient way to 
perform certain higher order perturbative calculations with path integrals. 


\section{A path integral derivation of complex WKB}

The aim of this section is to show how complex WKB emerges from
the path integral formulation of quantum mechanics. In Feynman's path 
integral formulation the wave function (we work for 
now in one space dimension) is written 
\begin{equation}
\label{path1}
\psi(X,T)=\int_{-\infty}^{\infty}K(x_0,X,T)\psi_0(x_0)dx_0,
\end{equation}
where $\psi_0(x_0)=\psi(x_0,0)$ is the initial wavefunction and
\begin{equation}
\label{path2} K(x_0,X,T)=\int {\cal{D}}x
\exp{\left(\frac{iS[x]}{\hbar}\right)}
\end{equation}
is the propagator. The propagator is represented as a sum over 
all possible paths $x(t)$, $0\leq t \leq T$, satisfying the 
boundary conditions $x(0)=x_0$ and $x(T)=X$. 
$S[x]$ denotes the classical action of the path $x$, given by 
\begin{equation}
S[x]=\int_{0}^{T} \frac12 m \dot{x}^{2}(t)-V(x(t))  \  dt .
\end{equation}
Inserting eq.(\ref{path2}) into eq.(\ref{path1}), moving
$\psi_0(x_0)$ into the argument of the exponent and incorporating
the integration over $x_0$ into $\int {\cal{D}}x$ yields an alternate
version of the path integral formulation
\begin{equation}
\label{path3} \psi(X,T)
=\int {\cal{D}}x  \exp{\left(\frac{iS[x]}{\hbar}+\log \psi_0(x(0)) \right)}
=\int {\cal{D}}x  
    \exp\left(\frac{i\left(S[x]+S^{\rm init}(x(0))\right)}{\hbar}\right) 
    \ ,
\end{equation}
where now $\int {\cal{D}}x$ represents now a sum over all possible
paths satisfying the  single boundary condition $x(T)=X$, and 
we have written, as before, $S^{\rm init}(x) = -i \hbar \log \psi_0(x)$.  

The next step is to evaluate
$\psi(X,T)$ using a saddle point approximation. To this end we
consider the variation of the term in the exponential in the path 
integral, and in particular identify paths for which the first order variation
vanishes. Replacing $x$ by $x+\epsilon$ in the term in the exponential 
we have 
\bea
S[x+\epsilon]+S^{\rm init}(x(0)+\epsilon(0)) 
&= &\int_0^{T} \frac12 m \left( \dot{x}+\dot{\epsilon} \right)^2
    -V(x(t)+\epsilon(t))  \  dt +S^{\rm init}(x(0)+\epsilon(0))\   ,
   \nonumber\\
&=& S[x]+S^{\rm init}(x(0))  + 
   \int_0^{T} \left( m\dot{x}\dot{\epsilon} - V'(x)\epsilon \right) dt 
    + {S^{\rm init}}'(x(0))\epsilon(0) 
    \nonumber \\
&& + \int_0^T \left(\frac12 m \dot{\epsilon}^2 - \frac12V''(x)\epsilon^2  \right)  dt 
    + \frac12{S^{\rm init}}''(x(0))\epsilon(0)^2  \nonumber\\
&& + \sum_{n=3}^\infty \frac1{n!} \left( {S^{\rm init}}^{(n)}(x(0))\epsilon(0)^n 
    - \int_0^T V^{(n)}(x)\epsilon^n   dt \right) \ .
\label{expan}\eea
After an integration by parts, and using the fact that $\epsilon(T)=0$, as 
all paths have the same fixed end point, the linear terms in $\epsilon$
become
\be - \int_0^{T} \left( m\ddot{x}  +  V'(x)\right) \epsilon dt 
    + \left( {S^{\rm init}}'(x(0))-m\dot{x}(0) \right) \epsilon(0) \ .\ee
Thus we deduce that {\em in a saddle point approximation of (\ref{path3}), the 
approximation will be a sum of contributions from classical paths 
satisfying the initial condition} 
\be \dot{x}(0) = \frac1{m} {S^{\rm init}}'(x(0)) = -\frac{i\hbar}{m} 
  \frac{\psi_0'(x(0))}{\psi_0(x(0))}\ .
\label{inc} \ee
These are  exactly the complex classical paths that appear in complex WKB.

Proceeding to look at the quadratic terms in (\ref{expan}), we want the variable 
over which we integrate in the path integral to be dimensionless, so we 
rescale $\epsilon$ by writing 
\be \epsilon(t) = \sqrt{\frac{\hbar T}{m}}\delta(t)\ . \ee
After this change the quadratic terms in (\ref{expan}) become 
\be \hbar T \left(
\int_0^T \left(\frac12 \dot{\delta}^2 - \frac1{2m}V''(x(t))\delta^2  \right)  dt 
    +\frac1{2m}{S^{\rm init}}''(x(0))\delta(0)^2  \right) \ . 
\ee
We are now in a position to write the saddle-point approximation to
(\ref{path3}):
\bea
\psi(X,T) &=& \sum_{x(t)} 
\exp\left(\frac{i\left(S[x]+S^{\rm init}(x(0))\right)}{\hbar}\right)  \label{pathsum}\\
&& \int {\cal{D}}\delta 
\exp\left( 
   i T \left(
\int_0^T \left(\frac12 \dot{\delta}^2(t) - \frac1{2m}V''(x(t))\delta^2(t)  \right)  dt 
    +\frac{{S^{\rm init}}''(x(0))}{2m}\delta(0)^2  \right)   \right) \nonumber \\
&& \exp
 \left( i 
  \sum_{n=3}^\infty \frac{\hbar^{\frac{n}{2}-1}}{n!}  
  \left(\frac{T}{m}\right)^{\frac{n}{2}} 
  \left( {S^{\rm init}}^{(n)}(x(0))\delta(0)^n - \int_0^T V^{(n)}(x)\delta^n(t) dt \right) 
   \right) \ .  \nonumber 
\eea
Here the sum is over complex WKB paths, that is paths $x(t)$ obeying the classical
equations of motion and the initial condition (\ref{inc}). However, as is usual in saddle-point
approximations, more detailed calculations are necessary to decide which of these
paths should be included in the sum. We will return to this point shortly. 

\subsection{The lowest order approximation} 

To compute the lowest order approximation we just need to evaluate the Gaussian 
integral
\be
\int {\cal{D}}\delta 
\exp\left( 
   i T \left(
\int_0^T \left(\frac12 \dot{\delta}^2(t) - \frac1{2m}V''(x(t))\delta^2(t)  \right)  dt 
    +\frac{{S^{\rm init}}''(x(0))}{2m}\delta(0)^2  \right)   \right) \ .
\ee
We recall that the integration here is over paths $\delta(t)$ obeying the single 
condition $\delta(T)=0$. As usual, we compute this integral by dividing the interval
$[0,T]$ into $N$ subintervals and discretizing. Appropriate (second order)
discretization formulas for the various terms are 
\bea 
\int_0^T \dot{\delta}^2(t)   dt  &\approx&
    \frac{N}{T}\left( \delta_0^2 + 2 \sum_{i=1}^{N-1}\delta_i^2 
       - 2 \sum_{i=0}^{N-1}\delta_i\delta_{i+1} \right)\ , \\
\int_0^T  V''(x(t))\delta^2(t)  dt &\approx&
 \frac{T}{N}\left( \frac12 V''(x(0))\delta_0^2 + 
\sum_{i=1}^{N-1} V''\left(x\left(\frac{iT}{N}\right)\right)\delta_{i}^2 
\right)\ ,
\eea 
where $\delta_i$  denotes  $\delta(iT/N)$. Using these, the discretized 
version of the path integral is 
\be
\int d^N\Delta \exp\left( \frac{iN}2 {\Delta} A {\Delta}^T \right) 
\label{basicint}\ee 
where 
$\Delta=\pmatrix{\delta_0 & \delta_1 & \delta_2 & \ldots & \delta_{N-1}}$, 
$A$ denotes the tridiagonal $N\times N$ matrix 
\be
A= \left(
\begin{array}{ccccc} 
q& -1                            & 0  & 0 & \ldots  \\
-1 & 2 - \frac{T^2}{mN^2}V''\left(x\left(\frac{T}{N}\right)\right) & -1 & 0 &   \ldots \\
0 & -1 & 2 - \frac{T^2}{mN^2}V''\left(x\left(\frac{2T}{N}\right)\right) & -1 & \ldots \\
0 & 0 & -1 & 2 - \frac{T^2}{mN^2}V''\left(x\left(\frac{3T}{N}\right)\right) &  \ldots  \\
\vdots & \vdots & \vdots & \vdots &  \\
\end{array}
\right) 
\ee
and 
\be
q = 1 - \frac{T^2}{2mN^2}V''(x(0)) + \frac{T{S^{\rm init}}''(x(0))}{mN}\ .
\ee
(At this point it is maybe worthwhile noting that for an initial Gaussian wavefunction, 
${S^{\rm init}}''(x(0))$ has a positive imaginary part.) The measure $d^N\Delta$ here 
includes a nontrivial $N$-dependent normalization; it turns out this should be chosen so
that  
\be
\int d^N\Delta \exp\left( \frac{iN}2 {\Delta} A {\Delta}^T \right)  = 
\frac1{\sqrt{\det A}} \ . 
\ee 
(For the standard rules for Gaussian integrals see for example \cite{wiki}; the 
correct choice of normalization is determined by checking that we get the 
correct result for a free particle.) 
The computation of the determinant $\det A$ proceeds as 
follows \cite{gelfand}: For $n=1,2,\ldots,N$, denote the determinant of the 
$n\times n$ matrix in the top left corner of $A$ by $D_n$. Then we have
\bea 
D_1 &=& q ~=~ 1 + \frac{T{S^{\rm init}}''(x(0))}{mN} + O\left( N^{-2} \right)  \  , \\
D_2 &=& q \left( 2 - \frac{T^2}{mN^2}V''\left(x\left(\frac{T}{N}\right)\right) \right) - 1
  ~=~ 1 + \frac{2T{S^{\rm init}}''(x(0))}{mN} + O\left( N^{-2} \right)  \  ,
\eea 
and for $3\le n\le N$
\be
D_n = \left( 2 - \frac{T^2}{mN^2}V''\left(x\left(\frac{(n-1)T}{N}\right)\right)\right) 
      D_{n-1} - D_{n-2} \ .
\ee 
The recursion can be written in the equivalent form 
\be \frac{D_n -2D_{n-1} + D_{n-2}}{(T/N)^2} =  
         - \frac1{m}V''\left(x\left(\frac{(n-1)T}{N}\right)\right) 
      D_{n-1} \ .
\label{Deq}\ee 
We need to determine $\det A=D_N$. The recursion and the initial conditions are such that 
as $N\rightarrow\infty$, the $D_n$ will tend to samples of a function $D(s)$, 
defined on the interval $0\le s\le T$, obeying the differential equation
$\ddot{D}(s)=-\frac1{m}V''(x(s))D$ and initial conditions $D(0)=1$ and $\dot{D}(0)=
\frac{1}{m} {S^{\rm init}}''(x(0))$. The determinant we seek is simply $\det A=D(T)$. 

To summarize, 
we have arrived at the lowest order approximation for the contribution of the path 
$x(t)$ in the sum (\ref{pathsum}): It is given by 
\be
\frac1{\sqrt{D(T)}}
\exp\left(\frac{i\left(S[x]+S^{\rm init}(x(0))\right)}{\hbar}\right)  \ .
\label{pathcont}\ee
Here $S[x]$ denotes the classical action associated with the path $x(t)$, which 
is a solution of Newton's equations obeying the conditions $x(T)=X$ and 
$\dot{x}(0) = \frac1{m} {S^{\rm init}}'(x(0))$. The function $S^{\rm init}$ is 
determined by the initial wave function via $S^{\rm init}(x)=-i\hbar\log\psi(x,0)$. 
The function $D(s)$ is the solution of the initial value problem 
\be
\ddot{D}(s)=-\frac1{m}V''(x(s))D\ , \qquad
D(0)=1\ ,~ \dot{D}(0)=\frac{1}{m} {S^{\rm init}}''(x(0))\ . 
\ee

We check that the above gives an exact result in the 
case of the free particle ($V=0$) and initial Gaussian wave function 
\be 
\psi(x,0) =  
\exp\left(
-\frac{a(x-x_0)^2}{\hbar} + \frac{ip_0(x-x_0)}{\hbar}
\right)\  . 
\ee
The initial wave function here has three parameters, $x_0$ and $p_0$ which are 
real and $a$ which is complex, with positive real part. Classical paths take the 
form $x(t)=A+Bt$. The coefficients $A,B$ should be determined by 
requiring 
\bea X=A+BT\ , \qquad B=\frac1{m}\left(p_0+2ia(A-x_0) \right)\ . \eea
The classical action along the path $x(t)$ is then given by $S[x]=\frac12 mB^2T$
and $D(T)=1+\frac{1}{m} {S^{\rm init}}''(x(0))T=1+\frac{2iaT}{m}$. 
Putting everything together we obtain
\bea
\psi(X,T) &=& 
\frac1{\sqrt{1+\frac{2iaT}{m}}}   
\exp\left( -\frac{a(A-x_0)^2}{\hbar} + \frac{ip_0(A-x_0)}{\hbar} \right)   
\exp\left( \frac{imB^2T}{2\hbar}  \right) \\
&=& \frac1{\sqrt{1+\frac{2iaT}{m}}}   
\exp\left(
-\frac{a}{\hbar\left(1+\frac{2iaT}{m}\right)}\left(X-x_0-\frac{p_0T}{m}\right)^2 
+\frac{ip_0}{\hbar}  \left(X-x_0-\frac{p_0T}{m}\right) 
+\frac{ip_0^2T}{2\hbar m}
\right)\ . \nonumber
\eea
It is straightforward to check that this is the exact solution of the Schr\"odinger
equation for the given initial condition. 

Having computed the lowest order approximation to the path integral in the one 
dimensional
case, we now state the generalization to the multidimemsional case, leaving 
the proof to an appendix. 
Once again the saddle point paths are exactly the 
trajectories that appeared in the complex WKB method, specifcially they are  classical 
paths ${\bf x}(t)$ obeying the initial condition
\be \dot{\bf x}(0) = \frac1{m} \nabla S^{\rm init}({\bf x}(0)) \ ,
\label{veq2}\ee
c.f. (\ref{veq}), as well as the final
condition ${\bf x}(T)={\bf X}$. The contribution from any such path to the wave
function $\psi({\bf X},T)$ takes the form 
\be
\frac1{\sqrt{D(T)}}
\exp\left(\frac{i\left(S[{\bf x}]+S^{\rm init}({\bf x}(0))\right)}{\hbar}\right)  \ .
\label{pathcont2}\ee
Here, as in the one-dimensional case, $S[{\bf x}]$ denotes the classicial action 
associated with the path ${\bf x}(t)$. The factor $D(T)$ is determined as follows: 
Denote by $U(s)$ the $d\times d$ matrix solution of the initial value problem
\be
\ddot{U}(s)=-\frac1{m}H(V)({\bf x}(s))U\ , \qquad
U(0)=I\ ,~ \dot{U}(0)=\frac{1}{m} H({S^{\rm init}})({\bf x}(0))\ ,
\label{Umx}\ee
where here $H(V)$ and $H(S^{\rm init})$ denote the $d\times d$ matrices 
of second derivatives of $V$ and ${S^{\rm init}}$ respectively. Then 
$D(T)=\det(U(T))$.  

We now wish to compare the lowest order path integral results with the 
lowest order approximation in complex WKB in the previous section. 
The path integral results all appear in the paragraph above. For ease, we
assemble here all the necessary equations from complex WKB.
The trajectories are determined from 
\be 
\frac{d{\bf x}}{dt} = {\bf v}\ ,\qquad
\frac{d{\bf v}}{dt} =  -\frac1{m} \nabla V({\bf x}) \ ,
\ee 
with boundary conditions 
\be {\bf v}(0) = \frac1{m} \nabla S_0^{\rm init}({\bf x}(0))\ , 
\qquad 
{\bf }{x}(T)={\bf X}\ .  
\ee 
The wave function is given by 
\be \psi({\bf X},T)=\exp\left(\frac{iS_0(T)}{\hbar}+S_1(T)\right)\ .  \ee 
The evolution equations of the necessary quantities along the trajectories are
\bea 
\frac{dS_0}{dt} &=& \frac12 m {\bf v}^2 - V({\bf x})\ ,   \\ 
\frac{dS_1}{dt} &=& \frac{i}{2m} \sum_{i=1}^d S_{0,ii}\ ,  \label{s1eq}\\
\frac{dS_{0,ik}}{dt}  &=& - V_{ik}\left({\bf x}(t)\right) -  
               \frac1{m} \sum_j S_{0,ij} S_{0,jk}\ ,  \label{mat2ders}
\eea 
with initial conditions
\be
S_0(0) =  S^{\rm init}({\bf x}(0))\ , \quad
S_{0,ij}(0)  =  S_{ij}^{\rm init}({\bf x}(0)) \ ,\quad
S_1(0)  =  0\ .  
\label{inicon}\ee
The correspondence 
is almost immediate. All that is necessary to do is to  identify the matrix 
with entries $S_{0,ij}$ in complex WKB with the matrix product 
$m\dot{U}U^{-1}$ in the 
path integral approach.  With this identification, the 
evolution equation  (\ref{mat2ders}) coincides with 
the second order evolution equation (\ref{Umx}) for $U$. 
Also after this identification, the evolution equation for $S_1$, (\ref{s1eq})
reads $\frac{dS_1}{dt} = \frac{i}{2} {\rm Tr}(\dot{U}U^{-1})$, with solution 
(taking into account the appropriate initial conditions) 
$S_1(t)=\frac{i}{2}\log \det U(t)$, so $e^{iS_1}=1/\sqrt{D(T)}$, giving the  
prefactor in (\ref{pathcont2}). Finally, 
$S_0$ in complex WKB is identified
with $S[{\bf x}]+S^{\rm init}({\bf x}(0))$ in the path integral approach. 

The path integral approach has added one significant piece of information over 
the direct complex WKB approach presented in the previous section. In the 
path integral approach we  use the saddle point method for asymptotic 
evaluation of an integral. As is well known, when there are multiple saddle 
points, it is sometimes necessary to take more than one into account 
to get an accurate approximation of the integral being studied. Deciding which
saddle points contribute requires detailed analysis on a case-to-case basis. But
at least we have found an explanation for the observations of our earlier work 
\cite{bomca2},\cite{cwkb} that for certain values of ${\bf X}$ and $T$ it 
is necessary to include the contibutions of multiple trajectories. (It is interesting
to compare this explanation for the origin of multiple trajectories with that given
by Miller \cite{Miller}, based on the implicit nature of the equations that generate dynamical
canonical transformations.  We suspect that Miller's explanation may correlate with
the existence of multiple solutions of the classical HJE, a connection that
would bring us full circle to an understanding of the need for multiple trajectories
in Complex WKB and BOMCA.)  In future
work \cite{caus} we hope to study the possible criteria for demarking different 
regions in ${\bf X},T$ space in which different (numbers of) trajectories 
contribute. This is strongly interconnected with 
the existence of {\em caustics}. Caustics are points ${\bf X},T$ at which 
the determinant $D(T)$ vanishes (on at least one trajectory, in fact such points 
are associated with coallescing trajectories). It is possible to study the 
dynamics of such points, and from this to deduce certain information about the 
dynamics of the regions in which different numbers of trajectories contribute. 
Unfortunately, however, at the moment deciding on which trajectories to include
in a calculation is more of an art than a science. 

\subsection{The first order correction} 

We now consider the higher order terms in (\ref{pathsum}). We restrict ourselves
in this section to the 1-dimensional case.  The series in the exponential
in the third line of (\ref{pathsum}) is an expansion in half-integer and integer 
powers of the 
dimensionless parameter $\frac{\hbar T}{m L^2}$ where $L$ denotes a typical 
length scale of the functions $V(x)$ and $S^{\rm init}(x)$. We are assuming 
this parameter is small. All terms with half-integer powers multiply odd powers
of $\delta$ and thus do not contribute to the value of the integral. The lowest
order correction terms arise when we replace the exponential by 
\bea
1  
&+& \frac{i \hbar  T^2}{24m^2}  
\left( {S^{\rm init}}''''(x(0))
\delta(0)^4 - \int_0^T V''''(x)\delta^4(t) dt \right) \nonumber \\
&-&  \frac{\hbar T^3}{72m^3}  
\left( {S^{\rm init}}'''(x(0))\delta(0)^3 - \int_0^T V'''(x)\delta^3(t) dt \right)^2 \ .
\eea
After discretizing, in this approximation the integral (\ref{basicint}) is replaced by 
an expression of the form
\be
\int d^N\Delta \exp\left( \frac{iN}2 {\Delta} A {\Delta}^T \right) 
\left( 1 + \sum_{i=0}^{N-1} \delta_i^4 {\cal A}_i 
+ \sum_{i=0}^{N-1}\sum_{j=0}^{N-1} \delta_i^3\delta_j^3 {\cal B}_{ij} 
\right) 
\ ,
\label{int2}\ee 
where ${\cal A}_i$ and ${\cal B}_{ij}$, which  do not depend on the components of $\delta$, are 
\bea
{\cal A}_{i} &=& \left\{ 
\begin{array}{ll}
\frac{i \hbar  T^2}{24m^2}  {S^{\rm init}}''''(x(0)) + O\left(\frac1{N}\right) &   i=0 
   \label{thecalA}\\
-\frac{i \hbar  T^3}{24m^2N} V''''\left(x\left(\frac{iT}{N}\right)\right) 
    &   i>0
\end{array}
\right. \ ,\\
{\cal B}_{ij} &=& \left\{ 
\begin{array}{ll}
- \frac{ \hbar  T^3}{72m^3} \left( {S^{\rm init}}'''(x(0)) \right)^2  + O\left(\frac1{N}\right) 
          &   i=j=0 \\
\frac{ \hbar  T^4}{72m^3N} {S^{\rm init}}'''(x(0)) V'''\left(x\left(\frac{iT}{N}\right)\right) 
    + O\left(\frac1{N^2}\right) 
    &   i>0,j=0 \\
\frac{ \hbar  T^4}{72m^3N} {S^{\rm init}}'''(x(0))
 V'''\left(x\left(\frac{jT}{N}\right)\right) 
    + O\left(\frac1{N^2}\right) 
    &   i=0,j>0 \\
- \frac{ \hbar  T^5}{72m^3N^2} V'''\left(x\left(\frac{iT}{N}\right)\right) 
                               V'''\left(x\left(\frac{jT}{N}\right)\right) 
    &   i,j>0  
\end{array}
\right. \ .
\eea
The Gaussian integrals in the above expression are standard. Taking into account 
our normalization of the measure $d^n\Delta$ we have 
\bea 
\int d^N\Delta \exp\left( \frac{iN}2 {\Delta} A {\Delta}^T \right) \delta_i^4  
&=& -   \frac3{N^2\sqrt{\det A}} \left( A^{-1}_{ii} \right)^2 \ , \label{delta4}\\
\int d^N\Delta \exp\left( \frac{iN}2 {\Delta} A {\Delta}^T \right) \delta_i^3 \delta_j^3 
&=&  -  \frac{3i}{N^3\sqrt{\det A}} \left(
3 A^{-1}_{ii} A^{-1}_{jj} A^{-1}_{ij} 
+ 2 \left( A^{-1}_{ij}  \right)^3  \right)  \label{delta6}
\eea
(c.f. \cite{wiki}). 
A calculation similar to the calculation of $\det A$ in the previous subsection
shows that as $N\rightarrow\infty$
\be \frac{T}{N} A^{-1}_{ij} \rightarrow  D(t_i)D(t_j) \int_{\max(t_i,t_j)}^T \frac{du}{D(u)^2}
  \ , \label{limit}\ee
where $D(s)$ is the solution of (\ref{Umx}). (This calculation uses the fact that a second, 
linearly independent, solution of the differential equation in (\ref{Umx}) is given by 
$D(s)\int_0^s\frac{du}{D(u)^2}$.) In this manner we can write down the first order
approximation to the path integral.
For simplicity  we restrict
ourselves here to the case that ${S^{\rm init}}'''$ and ${S^{\rm init}}''''$ 
vanish, as otherwise the relevant formulae are lengthy. Combining the formulae above 
we find that in this case the first order approximation is found by multiplying 
the leading order approximation by  
\bea
1 &+& \frac{i \hbar }{8m^2} 
\int_0^T
V''''\left(x(t)\right) D(t)^4 \left( \int_t^T \frac{du}{D(u)^2} \right)^2 \ dt\  
   \nonumber \\
&+&
\frac{i \hbar  }{24m^3} 
\int_{0}^T\int_{0}^T  
V'''\left(x(t_1)\right) 
V'''\left(x(t_2)\right)  
D(t_1)^3  D(t_2)^3 \label{foc1} \\
&& \left( 
3 \left( \int_{\max(t_1,t_2)}^T \frac{du}{D(u)^2} \right)
  \left( \int_{t_1}^T \frac{du}{D(u)^2} \right)
  \left( \int_{t_2}^T \frac{du}{D(u)^2} \right)
+ 
2 \left( \int_{\max(t_1,t_2)}^T \frac{du}{D(u)^2}  \right)^3  
\right)\ dt_1dt_2  
. \nonumber \eea

We now need to compare this with a similar term arising in the complex WKB method. 
The first order approximation in complex WKB is obtained by multiplying the leading 
order approximation by $e^{i\hbar S_2(T)}$. To find $S_2$ it is necessary to integrate
5 new differential equations along the trajectories (in addition to those that have to be 
solved to find the lowest order approximation): the equations for  
 $S_0''',S_0'''',S_1',S_1''$ and $S_2$, equations 
(\ref{mre2}),(\ref{mre3}),(\ref{S11}),(\ref{S12}) and
(\ref{S2e}) respectively.  All of these are linear equations, and an explicit  formula can 
be written for the answer. To simplify matters we assume the initial conditions 
for all the 5 relevant quantities are zero, which is consistent with the assumption
made in writing (\ref{foc1}). In its most obvious form 
(without making any attempts to simplify the integrals that appear) 
the solution takes the form 
\bea 
S_2(T) &=&  
\frac{1}{4m^2} \int_0^T \frac1{D^2(t)} 
 \left( \int_0^t \frac1{D^2(u)}\left( \int_0^u  V''''(x(v))D^4(v) dv  \right) du \right) dt  
\label{S2res}\\
&+& \frac1{8m^3} \int_0^T  \frac1{D^2(t)} 
\left( \int_0^t \frac1{D^2(u)} \left(\int_0^u V'''(x(v))D^3(v) dv\right)  du \right)^2 dt 
\nonumber\\
&+& \frac{3}{4m^3} \int_0^T \frac1{D^2(t)} 
\left( \int_0^t \frac1{D^2(u)} \left( \int_0^u \frac1{D^2(v)} 
\left( \int_0^v V'''(x(w))D^3(w) dw  \right)^2   dv \right) du \right) dt  
\nonumber\\
&+& \frac{1}{4m^3} \int_0^T \frac1{D^2(t)} \left( \int_0^t \frac1{D^2(u)} 
\left( \int_0^u V'''(x(v))D^3(v) dv  \right) 
\right. \nonumber\\
&& ~~~~~~~~~~~~~~~~~~~~~~~~~~~~~~~~~\left. 
\left( \int_0^u \frac1{D^2(v)}\left( \int_0^v V'''(x(w))D^3(w) dw  \right) dv \right)
~ du \right)  dt  \ . \nonumber
\eea
It is a straightforward but tedious matter to check that the factor (\ref{foc1}) 
is equal to $1+i\hbar S_2$ (the first order approximation to $e^{i\hbar S_2}$).

Thus we see explicitly the equivalence of the first order approximation to the 
path integral and results from the complex WKB method retaining terms up to 
order $S_2$. For consistency, this equivalence must continue to higher orders. Note 
that if we keep terms up to order $\hbar^n$ in the path integral the resulting formulae
will involve derivatives of $V$ (and $S^{\rm init}$) up to order $2n+2$, and the same
is true if we retain terms up to $S_{n+1}$ in complex WKB. We note that in practice, 
complex WKB is far easier to implement for higher order corrections. Although the 
number of differential equations that need to be integrated along the trajectories grows
rapidly with the order, as described in the previous section, it remains relatively 
easy to write down the necessary differential equations, and integrating the relevant 
first order system along the trajectories is easily handled using standard  computer
packages. Direct 
application of path integral methods involves the calculation of iterated integrals,
as in (\ref{S2res}) or (\ref{foc1}), which is a less standard procedure. The 
coefficients of the different iterated integrals (the number of which grows rapidly 
as order increases) also involve tricky combinatoric factors.

\section{A modification of standard asymptotic analysis}

In the previous section we have explained the connection of the complex WKB 
method as described in section 2 and the standard asymptotic evaluation of the 
path integral. We would like to also understand BOMCA  from this viewpoint. 
But there is a clear problem --- whereas the trajectories in complex WKB are
classical paths, corresponding to minima of the classical action, the paths
in BOMCA are nonclassical. How can nonclassical paths possibly arise in the context 
of an asymptotic evaluation of the path integral? In this section we describe a
modification of standard asymptotic analysis for Laplace-type
integrals. In the next section we will apply what we have learn here to path 
integrals. 

The usual approach to asymptotic evaluation of integrals such as 
$\int_{-\infty}^\infty g(x) e^{-\lambda f(x)} dx $, where $\lambda$ is 
a large positive parameter, proceeds as follows: The integral is 
dominated by contributions 
from regions close to the minima of $f(x)$. 
Sufficiently near a minimum $x_0$ 
the function $f(x)$ is approximated by a quadratic Taylor polynomial 
$f(x_0) + \frac12f''(x_0)(x-x_0)^2$. So we rewrite the integral in the form 
\be \int_{-\infty}^\infty \tilde{g}(x) e^{-\lambda \left( f(x_0)+
\frac12 f''(x_0)(x-x_0)^2\right)} dx\ ,  \ee
where $\tilde{g}(x)=g(x)\exp\left(-\lambda\left( f(x)-f(x_0) -
\frac12 f''(x_0)(x-x_0)^2\right)\right)$, and evaluate the contribution 
from the region near $x_0$ by expanding $\tilde{g}(x)$ in a Taylor series
in $x-x_0$ and evaluating the resulting integrals exactly. This gives a 
series in negative powers of $\lambda$. 

The modification to this procedure that we want to consider is as follows:
The Taylor polynomial approximation to $f(x)$ at its minimum is only one 
of many ways to approximate $f(x)$ in the appropriate region by a quadratic function. 
Suppose we choose another quadratic approximant. How does this change the 
resulting asymptotic expansion? 

For definiteness, we consider a specific example, asymptotic approximation 
of the factorial function for large $n$ using the integral 
representation
\be n! = \int_0^\infty e^{n\log x-x} dx\ . \label{fact}\ee
The function in the exponent has a minimum at $x=n$, and the usual 
asymptotic formula for $n!$ is obtained by approximating this function by 
the quadratic $n\log n-n - \frac1{2n}(x-n)^2$ and rewriting the integral 
\be n! \sim  n^ne^{-n} \int_{-\infty}^\infty e^{-(x-n)^2/2n} \tilde{g}(x)\ dx\ , \ee  
where 
\bea
\tilde{g}(x) &=& \exp\left( n\log x-x - (n\log n-n) + \frac1{2n}(x-n)^2 
  \right) \nonumber \\
&=& 1+ \frac1{3n^2}(x-n)^3-\frac1{4n^3}(x-n)^4+\frac1{5n^4}(x-n)^5
   +\frac{(n-3)}{18n^5}(x-n)^6 + \ldots  \label{tay}
\eea
Integrating gives the standard asymptotic series for $n!$
\be n!\sim \sqrt{2\pi}n^{n+\frac12}e^{-n}
\left(
1 + \frac1{12n} + \frac1{288n^2} - \frac{139}{51840n^3} + \ldots 
\right)\ .
\ee 
Note that to get the correct coefficient of $n^{-r}$ it is necessary to keep
certain terms of order up to $6r$ in the Taylor series (\ref{tay}). 

Suppose now that instead of using the above quadratic approximant for 
the exponent we use the more general approximant 
$ n \log N - N - \frac1{2S}(x-N)^2$. Here $S$ and $N$ are currently 
undetermined, but for definiteness we assume 
that $N=n+O(1)$ and $S=n+O(1)$. The integral now takes the form
\be
n! \sim N^n e^{-N} 
\int_{-\infty}^\infty 
 \exp\left(- \frac1{2S}(x-N)^2 \right)  \tilde{g}(x)  dx 
\ee 
where 
\bea
\tilde{g}(x) 
&=&
\exp\left(  n \log x - x - (n\log N - N) +\frac1{2S}(x-N)^2 
\right) \\
&=&  
\exp\left(  
\left(\frac{n}{N}-1\right)(x-N) + \frac12\left(\frac1{S}-\frac{n}{N^2}\right)(x-N)^2
+ \sum_{r=3}^\infty (-1)^{r-1}\frac{n}{rN^r}(x-N)^r 
\right) \ . \nonumber 
\eea
Making the substitution $x-N=\sqrt{S}y$ this becomes 
\be
n! \sim N^n e^{-N}\sqrt{S}  
\int_{-\infty}^\infty  e^{-y^2/2 }
\exp\left(  
\frac{(n-N)\sqrt{S}}{N} y  + \frac{N^2-nS}{2N^2} y^2
+ \sum_{r=3}^\infty (-1)^{r-1}\frac{nS^{r/2}}{rN^r}y^r 
\right) 
dy \ .
\label{bigfact}\ee 
Note that in the second exponent here the coefficients of $y$ and $y^3$ behave 
as $n^{-1/2}$,  the coefficients of $y^2$ and $y^4$ behaves as $n^{-1}$, and 
in general for
$r\ge 3$ the coefficient of $y^r$ behaves as $n^{1-r/2}$. We compute 
the integral by expanding the second exponential term in a power series
in $y$ and computing the resulting integrals exactly. The leading order
approximation is $ \sqrt{2\pi S}N^ne^{-N} $. The first order correction arises from 
replacing the second exponential by 
\be
1 +
\left(  \frac{N^2-nS}{2N^2} y^2   -  \frac{nS^2}{4N^4}y^4    \right) 
+  \frac12 
\left( \left(\frac{(n-N)\sqrt{S}}{N}\right) y  +   \frac{nS^{3/2}}{3N^3}y^3 \right)^2  \\
\ee
and integrating, to obtain 
\be n! \sim 
\sqrt{2\pi S}N^ne^{-N} \left( 1 + 
\frac1{12N^6} 
\left(
\begin{array}{c}
6N^6+10n^2S^3-6N^4nS+6N^4Sn^2-12N^5Sn +\\
6N^6S-9nS^2N^2+12n^2S^2N^2-12nS^2N^3
\end{array}
\right)
\right)  \ .  \label{modstir}
\ee 
It can be verified directly that provided $N=n+O(1)$ and $S=n+O(1)$ the correction 
term is $\frac1{12n} + O(n^{-2})$ for large $n$. Expanding the second exponential 
in (\ref{bigfact}) to suitable higher order gives us higher correction terms, apparently
depending on $N$ and $S$ as well as $n$, but in fact independent of the choice of $N,S$
to the desired order.  

Essentially what we have shown above is that Stirling's formula for $n!$ can be made
to depend on two variables $N$ and $S$ while retaining all its properties. The obvious 
question that needs to be asked at this stage is whether $N$ and $S$ can be chosen usefully. 
A full investigation of this would take us off on a tangent to the main topic of this paper,
so we limit ourselves here to the simple observation  which will allow us to 
give a path integral derivation of BOMCA: {\em It is possible to choose 
$N$ and $S$ as functions of $n$ in such a way that all correction terms to the 
leading order approximation $n!\sim \sqrt{2\pi S}N^ne^{-N}$ vanish.} Furthermore. at 
least in the case of the factorial function that we are looking at now, this is not
simply a perturbative result; that is, we can find analytic functions $N$ and $S$ of $n$, 
with the correct asymptotic behavior for large $|n|$ and such that $\Gamma(n+1) = 
\sqrt{2\pi S}N^ne^{-N}$ at least in some region of the complex plane including the 
positive real axis.  We will see in the next section how BOMCA is related to an 
analogous result for path integrals. Presumably there should be some way to select
$N$ and $S$ ``well'' on the basis of properties of the integrand of (\ref{fact}), but 
we do not attempt to study this here. 


\section{A derivation of BOMCA from the path integral} 

The path integral is 
\be
\psi(X,T)
= \int Dx  \exp \left( \frac{i}{\hbar} ( S[x]+S^{\rm init}(x(0)) ) \right) \ ,
\ee 
where as before the integration is over all paths with $x(T)=X$. Applying 
the idea presented in the previous section means approximating 
$S[x]+S^{\rm init}(x(0))$ with a quadratic, which we will take of the form
\bea
S[X]+S^{\rm init}(X(0)) &+& \int_0^T \frac12 m (\dot{x}(t)-\dot{X}(t))^2 - 
\frac12 \left( V''(X(t))+q(t) \right)(x(t)-X(t))^2 dt  \nonumber \\
&+& \frac12 ({S^{\rm init}}''(X(0))+q(0))(x(0)-X(0))^2 \ .
\eea
Here $X(t)$ is the path around which we are expanding, still to be 
fully determined, but assumed to 
be an order $\hbar$ perturbation of a classical path. Likewise the 
function $q(t)$ (which plays the role of $S$ in the previous section) 
is currently undetermined, assumed of order $\hbar$. Using this quadratic
as our leading order approximation in the path integral gives 
\bea 
\psi(X,T)
&=&  \exp \left( \frac{i}{\hbar} ( S[X]+S^{\rm init}(X(0)) ) \right)   
\int D\epsilon  \nonumber \\
&& \exp \left( \frac{i}{2\hbar} \left(
\int_0^T  m \dot{\epsilon}(t)^2 - 
 \left( V''(X(t))+q(t) \right)\epsilon(t)^2 dt  
+  ({S^{\rm init}}''(X(0))+q(0))\epsilon(0)^2 
 \right) \right)  \nonumber  \\
&& \exp \left( \frac{i}{\hbar} \left(
\int_0^T m\dot{X}(t)\dot{\epsilon}(t) - V'(X(t))\epsilon(t)+\frac12 q(t) \epsilon(t)^2 
- \sum_{r=3}^\infty \frac{V^{(r)}(X(t))}{r!}\epsilon(t)^r \ dt \right.\right. \nonumber  \\
&& \left.\left. + {S^{\rm init}}'(X(0))\epsilon(0) 
- q(0)\epsilon(0)^2  + 
\sum_{r=3}^\infty \frac{{S^{\rm init}}^{(r)}(X(0))}{r!}\epsilon(0)^r 
\right) \right)  \ . 
\eea
Here we have written $\epsilon(t)=x(t)-X(t)$. We can simplify the second 
exponential in the path integral in two ways. First, purely for 
ease of presentation we will assume that ${S^{\rm init}}^{(r)}=0$ for 
$r>3$, i.e. that the initial wave function is a Gaussian wave packet. There 
is no difficulty to restore the extra terms, but the calculations
become extremely lengthy. Second, we make choices on the initial values
of the currently unknown functions $X(t)$ and $q(t)$ to eliminate other 
terms in the second exponential as follows: First, we assume $q(0)=0$. Second, integrating 
the term $\int_0^T m\dot{X}(t)\dot{\epsilon}(t)$ gives a boundary contribution
$-m\dot{X}(0)\epsilon(0)$ and we can cancel this by requiring 
$m\dot{X}(0)={S^{\rm init}}'(X(0))$. Implementing all these simplifications gives us 
\bea 
\psi(X,T)
&=&  \exp \left( \frac{i}{\hbar} ( S[X]+S^{\rm init}(X(0)) ) \right)   
\int D\epsilon  \\
&& \exp \left( \frac{i}{2\hbar} \left(
\int_0^T m \dot{\epsilon}(t)^2 - 
 \left( V''(X(t))+q(t) \right)\epsilon(t)^2 dt  
+ {S^{\rm init}}''(X(0)) \epsilon(0)^2 
 \right) \right)  \nonumber  \\
&& \exp \left( \frac{i}{\hbar} \left(
\int_0^T (- m\ddot{X}(t) - V'(X(t)))\epsilon(t)+\frac12 q(t) \epsilon(t)^2 
- \sum_{r=3}^\infty \frac{V^{(r)}(X(t))}{r!}\epsilon(t)^r \ dt \right)\right) \ .\nonumber  
\eea
Finally, we move to dimensionless quantities by substituting 
$\epsilon(t)=\sqrt{\frac{hT}{m}}\delta(t)$, giving
\bea 
\psi(X,T)
&=&  \exp \left( \frac{i}{\hbar} ( S[X]+S^{\rm init}(X(0)) ) \right)   
\int D\delta  \nonumber \\
&& \exp \left(  \frac{iT}{2} \left(
\int_0^T \dot{\delta}(t)^2 - 
\frac1{m} \left( V''(X(t))+q(t) \right)\delta(t)^2 dt  
+ \frac1{m} {S^{\rm init}}''(X(0)) \delta(0)^2 
 \right) \right)  \nonumber  \\
&& \exp \left( i \left( \int_0^T 
   - \sqrt{\frac{mT}{\hbar}}\left(\ddot{X}(t) +\frac1{m} V'(X(t))\right)\delta(t)
   + \frac{T}{2m} q(t) \delta(t)^2  \right.\right.\nonumber\\
&& ~~~~~~~\left.\left. 
   - \sum_{r=3}^\infty \frac{\hbar^{r/2-1}T^{r/2}V^{(r)}(X(t))}{m^{r/2}r!}\delta(t)^r 
 dt \right)\right) \ .
\eea
Assuming that both $\ddot{X}(t)+\frac1{m}V'(X(t))$ and $q(t)$ are of order $\hbar$, we
see that the coefficients of $\delta(t)$ and $\delta(t)^3$ in the second exponential
are of order $\hbar^{-1/2}$, the coefficients of 
$\delta(t)^2$ and $\delta(t)^4$ are of order $\hbar^1$, and in general the coefficient
of $\delta(t)^r$ is of order $\hbar^{r/2-1}$ for $r\ge 3$. (This is in direct analogy
to the caclulations for the factorial function in section 4.) The leading order 
approximation to the 
path integral is obtained by simply discarding the second exponential term. The 
remaining Gaussian integral is identical to one we have already computed (but with 
$V''(X(t))$ replaced by $V''(X(t))+q(t)$), and we obtain the leading order
approximation 
\be
\psi(X,T) = \frac{\exp \left( \frac{i}{\hbar} ( S[X]+S^{\rm init}(X(0)) ) \right)}
       {\sqrt{f(T)}} 
\label{loabom}\ee 
where $f(s)$ is the solution of  
\be
\ddot{f}(s)=-\frac1{m}(V''(X(s))+q(s))f(s)\ , \qquad
f(0)=1\ ,~ \dot{f}(0)=\frac{1}{m} {S^{\rm init}}''(X(0))\ . 
\ee
To obtain the first order correction to this, we need to replace the second 
exponential in the path integral by 
\bea
1 
&+& 
\frac{i\hbar T}{2m} \left( \int_0^T 
 \frac{q(t)}{\hbar} \delta(t)^2    
- \frac{ TV''''(X(t))}{12m}\delta(t)^4 
\ dt \right)  \\
&-&
\frac{\hbar m T}{2}
\left( \int_0^T 
\frac{\left(\ddot{X}(t) +\frac1{m} V'(X(t))\right)}{\hbar}\delta(t)
+ \frac{TV'''(X(t))}{6m^2}\delta(t)^3 
\ dt \right)^2 \ .
\nonumber\eea
There are 5 terms here that we need to consider, as oppposed to 2 in the derivation of 
the first correction term in 
Complex WKB. In addition to the integration formulae (\ref{delta4})-(\ref{delta6}) 
we need the  formulae 
\bea 
\int d^N\Delta \exp\left( \frac{iN}2 {\Delta} A {\Delta}^T \right) \delta_i \delta_j
&=&  \frac{i}{N\sqrt{\det A}} A^{-1}_{ij} \ , \\
\int d^N\Delta \exp\left( \frac{iN}2 {\Delta} A {\Delta}^T \right) \delta_i^3 \delta_j 
&=&  -  \frac{3}{N^2\sqrt{\det A}} A^{-1}_{ii}  A^{-1}_{ij} \ .
\eea
Computing all the necessary integrals gives the following result for the first order
correction: The leading order approximation should be multiplied by  
\bea
1  &-& \frac1{2m}\int_0^T q(t)f(t)^2 \left( \int_t^T \frac{du}{f(u)^2} \right)\ dt \\
&-& \frac{im}{2\hbar} \int_0^T \int_0^T N(t_1)N(t_2)f(t_1)f(t_2) 
   \left( \int_{{\rm max}(t_1,t_2)}^T \frac{du}{f(u)^2}\right)\  dt_1 dt_2  \nonumber \\
&+& \frac{1}{2m} \int_0^T \int_0^T N(t_1)V'''(X(t_2))f(t_1)f(t_2)^3
    \left( \int_{{\rm max}(t_1,t_2)}^T \frac{du}{f(u)^2}\right)
    \left( \int_{t_2}^T \frac{du}{f(u)^2}\right)
    \  dt_1 dt_2  \nonumber \\
&+& \frac{i \hbar }{8m^2} 
\int_0^T
V''''\left(X(t)\right) f(t)^4 \left( \int_t^T \frac{du}{f(u)^2} \right)^2 \ dt\  
   \nonumber \\
&+&
\frac{i \hbar  }{24m^3} 
\int_{0}^T\int_{0}^T  
V'''\left(X(t_1)\right) 
V'''\left(X(t_2)\right)  
f(t_1)^3  f(t_2)^3   \nonumber \\
&& \left( 
3 \left( \int_{\max(t_1,t_2)}^T \frac{du}{f(u)^2} \right)
  \left( \int_{t_1}^T \frac{du}{f(u)^2} \right)
  \left( \int_{t_2}^T \frac{du}{f(u)^2} \right)
+ 
2 \left( \int_{\max(t_1,t_2)}^T \frac{du}{f(u)^2}  \right)^3  
\right)\ dt_1dt_2  
. \nonumber \eea
Here we have written $N(t)=\ddot{X}(t)+\frac1{m}V'(X(t))$. We have
left the integrals here in the form they arise using the relevant rules for Gaussian
integrals. To manipulate the integrals, though, it is more convenient to 
write them in terms of 
integrals in which all the variables are all ordered. Doing this gives: 
\bea 1
&-& \frac{1}{2m} \int_0^T dt_1 \int_{t_1}^{T} dt_2  \ \ 
q(t_1) f(t_1)^2 \frac1{f(t_2)^2}  \\ 
&-&  \frac{im}{\hbar }\int_0^T dt_1 \int_{t_1}^{T} dt_2 \int_{t_2}^{T} dt_3   \ \ 
N(t_1)f(t_1) N(t_2)f(t_2) \frac1{f(t_3)^2}
\nonumber \\ 
&+&  \frac{1}{m} \int_0^T dt_1 \int_{t_1}^{T} dt_2 \int_{t_2}^{T} dt_3 \int_{t_3}^{T} dt_4   \ \ 
N(t_1)f(t_1) V'''(t_2)f(t_2)^3   \frac1{f(t_3)^2}  \frac1{f(t_4)^2}
\nonumber \\ 
&+&  \frac{1}{2m} \int_0^T dt_1 \int_{t_1}^{T} dt_2 \int_{t_2}^{T} dt_3 \int_{t_3}^{T} dt_4   \ \ 
V'''(t_1)f(t_1)^3  \frac1{f(t_2)^2} N(t_3)f(t_3)  \frac1{f(t_4)^2}
\nonumber \\ 
&+&  \frac{1}{m} \int_0^T dt_1 \int_{t_1}^{T} dt_2 \int_{t_2}^{T} dt_3 \int_{t_3}^{T} dt_4   \ \ 
V'''(t_1)f(t_1)^3  N(t_2)f(t_2) \frac1{f(t_3)^2}  \frac1{f(t_4)^2}
\nonumber \\ 
&+&  \frac{i\hbar}{4m^2} \int_0^T dt_1 \int_{t_1}^{T} dt_2 \int_{t_2}^{T} dt_3   \ \ 
V''''(t_1)f(t_1)^4  \frac1{f(t_2)^2} \frac1{f(t_3)^2}
\nonumber \\ 
&+&  \frac{i\hbar}{2m^3}
\int_0^T dt_1 \int_{t_1}^{T} dt_2 \int_{t_2}^{T} dt_3 \int_{t_3}^{T} dt_4 \int_{t_4}^T dt_5 \ \ 
V'''(t_1)f(t_1)^3   \frac1{f(t_2)^2} V'''(t_3)f(t_3)^3  \frac1{f(t_4)^2} \frac1{f(t_5)^2}
\nonumber \\ 
&+&  \frac{5 i\hbar}{2m^3}
\int_0^T dt_1 \int_{t_1}^{T} dt_2 \int_{t_2}^{T} dt_3 \int_{t_3}^{T} dt_4 \int_{t_4}^T dt_5 \ \ 
V'''(t_1)f(t_1)^3   V'''(t_2)f(t_2)^3 \frac1{f(t_3)^2} \frac1{f(t_4)^2} \frac1{f(t_5)^2}
\nonumber 
\eea 
Our intention now is to choose the functions $N(t)$ and $q(t)$ (both assumed to be 
of order $\hbar$) in such a way that there is no first order correction, i.e. so that
the sum of the integrals in the above expression vanishes. From the above we see immediately
that {\sl for any choice of $N(t)$ it is possible to choose $q(t)$ such that the 
first order correction terms vanish}. One choice that suggests itself for $N(t)$ is 
simply to take $N(t)=0$. Then the correct choice of $q(t)$ is 
\bea
q(t)&=& \frac{i\hbar}{mf(t)^4}\left( 
  \frac12 \int_0^t du f(u)^4V''''(X(u)) \right. \label{otherq}\\
&& + 
\frac1{m} \int_0^t du \int_0^u dv \int_0^v dw V'''(X(u))f(u)^3 \frac1{f(v)^2} V'''(X(w))f(w)^3 
\nonumber \\
&& \left.
+ \frac5{m} \int_0^t du \int_0^u dv \int_0^v dw \frac1{f(u)^2} V'''(X(v))f(v)^3 V'''(X(w))f(w)^3 
\right)\ . \nonumber 
\eea 
The solution that is of main interest for us, however, is 
\bea
N(t) &=& - \frac{i\hbar}{2m^2f(t)^3}\int_0^t V'''(X(u)) f(u)^3 du 
  \label{N1}\\
q(t) &=& \frac{i\hbar}{mf(t)^4}\left( 
  \frac12 \int_0^t du f(u)^4V''''(X(u)) \right. \label{q1}\\
&& \left.
+ \frac3{m} \int_0^t du \int_0^u dv \int_0^v dw \frac1{f(u)^2} V'''(X(v))f(v)^3 V'''(X(w))f(w)^3 
\right) \nonumber 
\eea 

We summarize what we have shown up to this point. For either of the choices of 
$N(t)$ and $q(t)$ given above (or for any other choice of $N(t)$ and the appropriate 
matching choice of $q(t)$) we have demonstrated that the leading order approximation 
to the path integral,  (\ref{loabom}), requires no first order correction. Here the path 
$X$ and the function $f$ are chosen to satisfy 
\bea 
&&\ddot{X}(t) + \frac1{m} V'(X(t)) = N(t)\ , ~~
m\dot{X}(0)={S^{\rm init}}'(X(0))\ , ~X(T)=X \ ,  \label{bo1}\\
&& m \ddot{f}(s) + V''(X(s))f(s) = -q(s)f(s)\ , ~~
f(0)=1\ ,~ \dot{f}(0)=\frac{1}{m} {S^{\rm init}}''(X(0))\ . \label{bo22}
\eea

It is straightforward to check that the choice (\ref{N1})-(\ref{q1}) describes BOMCA (compare
equations (\ref{loabom}),(\ref{N1}),(\ref{q1}),(\ref{bo1}),(\ref{bo22}) 
with (\ref{zzz1}),(\ref{zzz2}),(\ref{zzz3}),(\ref{zzz4});
the constants $K,L$ should be chosen to be zero, and recall that in the discussion
of BOMCA we wrote $S''=\frac{m}{f}\frac{df}{dt}$). We have arrived at the understanding 
of BOMCA set out in the introduction --- that it corresponds to  an evaluation of the 
path integral around a near-classical path, chosen in such a way that the classical 
wave function remains accurate to any desired order in $\hbar$, with the path 
${\bf x}$ and $U$ being modified
appropriately. We have found that in fact there are other ways to change ${\bf x}$ and $U$
in such a way as to ``correct'' the classical
wave function. In particular, we can continue to use classical paths, but replace the usual
Jacobi equation  with (\ref{bo22}), where 
$q$ is given by  (\ref{otherq}). (The existence of this option extends to higher dimensions.)
In practice the direct derivation of BOMCA, as given in section 2,  
is clearly preferable over the path integral for determining higher order corrections. 
The path integral approach, however, is necessary to understand the 
need to add contributions from different (near-)classical
trajectories can be justified. 

Both the direct approach to BOMCA and the path integral approach only allow
us to construct the relevant near-classical trajectories order-by-order in $\hbar$. 
The question arises as to whether it is possible to find pairs ${\bf x}$ and $U$ for which 
the classical wave function is exact. As we have already explained in the introduction, 
the relevant paths would be an intermediate object between classical paths and the 
quantum trajectories of Bohmian mechanics --- on the one hand the new paths would 
be order $\hbar$ perturbations of classical paths, but on the other hand they would 
enable permit the derivation of exact quantum dynamical results, at least in regions 
of configuration space where they exist. At this stage the existence of such paths
remains just a conjecture. 


\section{The coherent state propagator}

This section is a slight digression from the main point of this paper, but 
provides another illustration of our path integral methods, as well as the need 
for  complex classical trajectories in ``semiclassical'' calculations. 
The so-called coherent state propagator
has been studied extensively by many authors 
\cite{Klauder1,Klauder2,Weissman,Adachi,XdA1,XdA2,XdA3,VVH1,VVH2,deA1,deA2}. 
By the coherent state propagator 
we mean the overlap between the wave function $\psi({\bf x},T)$ evolving from
an (initial) coherent state with another (final) coherent state. In fact our methods
allow us to write down a rather more general object; we write down the leading 
order approximation for the overlap 
between the wave function $\psi({\bf x},T)$ evolving from
an initial state of form $\exp(iS_i({\bf x})/\hbar)$ 
with a final state of the form $\exp(iS_f({\bf x})/\hbar)$. Using the 
Feynmann path integral representation of the (standard) propagator, the 
overlap  takes the form 
\be 
{\cal P} = 
\int_{-\infty}^\infty d{\bf x}_f  \psi_f^*({\bf x}_f)
\int_{-\infty}^\infty d{\bf x}_i  \psi_i({\bf x}_i)
\int {\cal D}{\bf x} \ \exp\left( \frac{i S[{\bf x}]}{\hbar}\right) 
\ee 
where the path integral is over all paths satisfying ${\bf x}(0)={\bf x}_i$,
${\bf x}(T)={\bf x}_f$. We can absorb the integrations
over the initial and final position into the path integral to write this 
simply as 
\be 
{\cal P} = 
\int {\cal D}{\bf x} \ 
\exp\left(\frac{i \left( S[{\bf x}] + S_i({\bf x}(0)) - S_f^*({\bf x}(T)) \right)}{\hbar}\right) 
\ee
where now the path integration is over all paths ${\bf x}(t)$, $0\le t \le T$, 
with no specified boundary conditions. 

To compute the semiclassical approximation to ${\cal P}$ we replace ${\bf x}$ in 
the exponent in the above expression  by ${\bf x}+\ve$ and expand to second order
in $\ve$. We choose ${\bf x}$ so that the linear term vanishes. Taking the action
to be given by $\int_0^T \frac12 m\dot{\bf x}^2 - V({\bf x})\  dt$ we find that the 
appropriate paths must satisfy 
\bea
m \ddot{\bf x} + \nabla V({\bf x}) &=& 0 \ , \\
m \dot{\bf x}(0) &=& \nabla S_i({\bf x}(0))\ , \\
m \dot{\bf x}(T) &=& \nabla S_f^*({\bf x}(T))\ .
\eea
The leading order approximation is thus a sum over such paths of the form 
\be 
{\cal P} \approx 
\sum \psi_f^*({\bf x}(T)) \exp(i S[{\bf x}]/\hbar) \psi_i({\bf x}(0)) 
  \int {\cal D}\ve \ \exp( Q[\ve] ) 
\ee
where 
\bea
Q[\ve] &=& \frac{i}{2\hbar} \Bigg( 
\int_0^T m \dot{\ve}^2 - \ve(t)^T H(V)({\bf x}(t)) \ve(t) \ dt 
 \nonumber \\
&&  + \ve(0)^T H(S_i)({\bf x}(0)) \ve(0) 
- \ve(T)^T H(S_f^*)({\bf x}(T)) \ve(T) 
\Bigg)\ . 
\eea 
Following  the method of the calculation in appendix A, the factor 
$  \int {\cal D}\ve \ \exp( Q[\ve] )$ 
can be replaced (modulo some normalization factor)  by 
the large $N$ limit of $1/\sqrt{\det V}$, where 
\be
V = 
\left( 
\begin{array}{ccccccc} 
I+G_0  & -I     & 0      & 0      &   0 & \ldots  &  \\
-I     & 2I+G_1 & -I      & 0      &   0 & \ldots  &   \\ 
0     & -I      & 2I+G_2 & -I      &   0 &         &  \\
0     & 0      & -I     & 2I+G_3 &   -I &         &  \\
\vdots & \vdots&        &   \ddots     &  \ddots   &      \\
       &       &        &        &     -I  & 2I+G_{N-1} & -I \\
        &      &        &        &         &     -I   & I+G_N 
\end{array}
\right) \ .
\ee
Here
\bea
G_0 &=&  \frac{T}{mN}H(S_i)({\bf x}(0)) - 
                \frac{T^2}{2mN^2}H(V)({\bf x}(0))\ ,  \\
G_r &=&  - \frac{T^2}{mN^2}H(V)\left({\bf x}\left(\frac{rT}{N}\right)\right) 
           \ ,\qquad r=1,\ldots,N-1 \ , \\
G_N &=&  - \frac{T}{mN}H(S_f^*)({\bf x}(T)) - 
                \frac{T^2}{2mN^2}H(V)({\bf x}(T)) \ . 
\eea
The method for evaluating the determinant used in appendix A (with a slight 
addition to understand the nontrivial normalization) 
yields the final result
\be 
{\cal P} \approx \left(\frac{2\pi i\hbar}{m} \right)^{d/2}
\sum \frac{\psi_f^*({\bf x}(T)) \exp(i S[{\bf x}]/\hbar) \psi_i({\bf x}(0))}
  {\sqrt{\det\left( 
\begin{array}{c} 
\dot{U}_1(T) + \frac1{m} \dot{U}_2(T) H(S_i)({\bf x}(0)) 
 - \frac1{m}  H(S_f^*)({\bf x}(T)) U_1(T) -  \\ 
\frac1{m^2}  H(S_f^*)({\bf x}(T)) U_2(T) H(S_i)({\bf x}(0)) 
\end{array}
     \right)}}
\ee
Here $U_1$ and $U_2$ are two solutions of the equation 
\be m\ddot{U}(t) = - H(V) U(t) \ , \ee
satisfying initial conditions 
\be 
\left\{ \begin{array}{ccc} U_1(0) & = & I \\ \dot{U}_1(0) & = & 0 \end{array}
\right. \ ,
\qquad\qquad
\left\{ \begin{array}{ccc} U_2(0) & = & 0 \\ \dot{U}_2(0) & = & I \end{array}\ .
\right.
\label{139}\ee
From equation (\ref{139}) it follows that 
the entries of $U_2$ have dimensions of time, whereas those of $U_1$ 
are dimensionless. 

Restricting to the case of 
Gaussian initial and final states, taken in the form 
\bea 
\psi_i({\bf x}) &=&  \exp\left(
  - \frac{m({\bf x}-{\bf x}_{0i})^T\Omega_i({\bf x}-{\bf x}_{0i})}{2\hbar} 
 + \frac{i{\bf p}_{0i}\cdot({\bf x}-{\bf x}_{0i})}{\hbar} \right)\ , \\
\psi_f({\bf x}) &=&  \exp\left(
  - \frac{m({\bf x}-{\bf x}_{0f})^T\Omega_f({\bf x}-{\bf x}_{0f})}{2\hbar} 
 + \frac{i{\bf p}_{0f}\cdot({\bf x}-{\bf x}_{0f})}{\hbar} \right)\ , 
\eea
the 
above formula reduces to 
\be 
{\cal P} \approx \left(\frac{2\pi i\hbar}{m} \right)^{d/2}
\sum \frac{\psi_f^*({\bf x}(T)) \exp(i S[{\bf x}]/\hbar) \psi_i({\bf x}(0))}
  {\sqrt{\det\left( 
 \dot{U}_1(T) + i \dot{U}_2(T) \Omega_i
 + i \Omega_f^* U_1(T) +  \Omega_f^* U_2(T) \Omega_i
     \right)} } \ . 
\label{Pform}\ee
Here ${\bf x}_{0i},{\bf p}_{0i}$ are real parameters giving the expectation 
values of 
the position and momentum in the intial state, 
${\bf x}_{0f},{\bf p}_{0f}$ are real parameters giving the expectation of 
the position and momentum in the final state, and $\Omega_i,\Omega_f$ are symmetric, 
complex matrices (with eigenvalues with positive real part). The relevant 
paths in the case of Gaussian initial states are those satisfying the boundary 
conditions
\bea 
m\dot{\bf x}(0)&=&{\bf p}_{0i}+ im\Omega_i({\bf x}(0)-{\bf x}_{0i})  \ , \label{bc0}\\
m\dot{\bf x}(T)&=&{\bf p}_{0f}- im\Omega_f^*({\bf x}(T)-{\bf x}_{0f})  \ , \label{bcT}
\eea
c.f. \cite{Klauder1,Klauder2,Weissman,Adachi,XdA1,XdA2,XdA3,VVH1,VVH2,deA1,deA2}. 
Note that the formula (\ref{Pform}) is not dimensionless as for simplicity we have 
been working with nonnormalized Gaussian states. 

We give one further
simplifcation, just as an illustration of the use of this formula: In the scalar 
case for a free particle ($U_1(t)=1,U_2(t)=t$) the formula gives the exact result
\bea
&&{\cal P} ~=~ \sqrt{\frac{2\pi \hbar}{m(\Omega_i+\Omega_f^*+iT\Omega_i\Omega_f^*)}} \\
&&\exp\left( - \frac
{(p_i-p_f)^2 + m^2\Omega_i\Omega_f^*(x_i-x_f)^2+iT(p_i^2\Omega_f^*+p_f^2\Omega_i)
+2im(x_i-x_f)(p_i\Omega_f^*+p_f\Omega_i)
}
{2\hbar m\left(\Omega_i+\Omega_f^*+iT\Omega_i\Omega_f^* \right) }
\right) \ .  \nonumber
\eea
(Here we have slightly changed notation, dropping the ``0'' suffices on the position
and momentum parameters.) 
It can be verified that 
the exponential is a pure phase if and only $p_f=p_i$, $x_f=x_i+p_it/m$, in which case
it becomes simply $\exp(itp_i^2/2m\hbar)$. 

The initial and final conditions on the complex trajectories (\ref{bc0})-(\ref{bcT}) are 
familiar from the literature, see  in particular \cite{deA1}. The 
semiclassical approximation (\ref{Pform}) is presented somewhat  differently from 
from formulae in the literature, but it would seem to be equivalent. Our 
derivation, while maybe not as careful as previous derivations, is a substantial
simplification. 

\section{Concluding remarks}

The main results of this paper are as follows: After a detailed presentation of
the complex WKB and BOMCA methods we showed how complex WKB can be derived from
a saddle point approximation in the path integral formulation of quantum mechanics.
The path integral approach to the method explains the need to incorporate 
the contributions from multiple trajectories; the original formulation however is 
much more useful for practical applications, avoiding the cumbersome multiple 
integrals that arise when computing higher order correction terms from the path
integral. In terms of methodology, the novel aspect of our path integral derivation
was incorporation of the initial wave function into the integrand prior to computing 
the saddle points and the relevant behavior near them: It is this that gives rise
to complex trajectories. Complex and real trajectory methods in quantum mechanics
are complementary, not contradictory --- complex trajectories are needed to propagate
wave packet-type states as considered in this paper, whereas real trajectories 
suffice for WKB-type states. 

We then moved on to the path integral description of BOMCA. This required a further
methodological innovation, the use of a general quadratic approximation in 
asymptotic analysis, as opposed to the standard Taylor approximation at the minimum. 
Using this more general asymptotic method we showed how to obtain BOMCA from the 
path integral (thus justifying the need for multiple trajectories in BOMCA too). In 
fact, from the path integral point of view at this stage BOMCA seems to be just one 
of many possible methods, a matter that merits further investigation. The overall picture 
of the relationship between complex WKB and BOMCA became clear. Both methods give 
rise to the same lowest order approximation to the wave function, the ``classical 
wave function'' (\ref{clwf}). In complex WKB this approximation is refined by multiplying
the wave function 
by suitable factors of the form $1+O(\hbar)$, while keeping the same trajectories 
${\bf x}$ and their variations $U$. In BOMCA, it is the formula (\ref{clwf}) that remains 
the same, while $O(\hbar)$ corrections are made to the trajectories ${\bf x}$ and the 
matrices $U$; these are changed depending on the order of the approximation. 

In section 6 we showed how our method of inserting the wave function into the path 
integral prior to making a saddle point approximation could be used to derive the 
coherent state propagator, measuring the overlap between an evolved Gaussian wave 
packet and another Gaussian state. Our derivation is a substantial simplifcation over
previous ones. In
the case of the coherent state propagator  there is no alternative 
derivation to the path integral; for computations of the wave function, 
however, we emphasize that the 
derivation of the equations of complex WKB and BOMCA presented in section 2 is 
simpler than the path integral approach, which is only necessary to explain 
the need for multiple trajectories. 

There are a number of areas in which further work is necessary. First and foremost, this
paper was intended to provide the theoretical backing for the numerical studies in 
\cite{bomca1,bomca2,cwkb}, and having done this, we hope that further numerical studies will
be undertaken, especially in multiple dimensions. There are several areas in which more
theoretical developments would be welcome. First, we have almost completely avoided in 
this paper any discussion of caustics (points at which the denominator in (\ref{clwf}) vanishes,
rendering the approximation meaningless) and the related phenomenon of Stokes' lines. 
In the case of wave function approximations, the caustics and Stokes' lines are 
dependent on the time, and it is possible to write down equations describing their motion
\cite{caus}. It is widely appreciated that the phenomena of caustics and Stokes' lines
are ``coordinate dependent'', in the sense that they can be avoided (or moved) by 
working in momentum or phase space representations \cite{regcau1,regcau2,regcau3,regcau4}. 
However, not enough has been
done yet to make these ideas into efficient techniques for calculations. Strongly 
related to these questions is the more mathematical question of the nature of the 
singularities in the system of ODEs arising in complex WKB at a caustic, which we are 
currently investigating \cite{jsprep}.  

Another matter requiring further investigation is a better understanding of the 
(linear) decomposition of the wave function  implied in (\ref{clwf}). Given an 
initial wave function is it possible (either abstractly or operationally) to
write it as a sum of terms each of which evolves into one of the terms in 
the sum (\ref{clwf})? Is the evolution by the Schr\"odinger equation or some other
equation? Many ideas in these directions have been discussed by Poirier 
and collaborators \cite{B1,B2,B3,B4,B5,B6}. 

Finally, we mention that we find the perturbed Newton's equations appearing in 
BOMCA to be fascinating. As mentioned above, from the path integral approach it
emerges that the BOMCA equations are not unique, and we would like a way to select
the version that emerges directly from the Schr\"odinger equation in section 2. 
We strongly suspect this to be related to some symmetry structure (recall that 
the underlying Newton's equations are Hamiltonian), but have not yet found 
this structure. Understanding this might 
give us clues as to how to find nonperturbative BOMCA trajectories, that is 
trajectories that when used in (\ref{clwf}) give exact answers. In addition to these
challenging, long-term goals, there is much to be done in seeking solutions 
of the first order BOMCA equations for specific systems and understanding, for example, 
the  difference between the behavior of complex WKB and BOMCA near caustics. 


\section*{Acknowledgements} J.S. wishes to thank David Kessler and Harry Dym for 
many useful discussions. Much of this work was done while J.S. was on sabbatical
leave in the Departments of Mathematics and Chemical Physics at the Weizmann Institute. 
This work was supported by the Israel Science Foundation (576/04). 


\section*{Appendix A: Derivation of the classical wave function in the 
multidimensional case}

In this appendix we derive the multidimensional form of the classical wave function 
as described in the introduction and in section 3, see equations 
(\ref{veq2}),(\ref{pathcont2}),(\ref{Umx}) and the text around them. The result differs 
slightly from standard results, specifically in the initial conditions obeyed
by the classical paths (\ref{veq2}) and the function $U$ (\ref{Umx}), and in any case the 
relevant calculations in the multidimensional case do not seem to have made it into 
most of the existing texts on path integration techniques, so we see fit to give 
at least the key details of the derivation. 

We start from the path integral in the form
$$
\psi({\bf X},T)
= \int D{\bf x}  \exp \left( \frac{i}{\hbar} ( S[{\bf x}]+S^{\rm init}({\bf x}(0)) ) 
    \right) \ , 
$$
where the integration is over all paths with ${\bf x}(T)={\bf X}$. 
We use, in this appendix, an action of the form 
$$
S[{\bf x}]=\int_0^T \frac12 \sum_{i=1}^d m_i \dot{x}_i(t)^2 - V({\bf x}(t)) \ dt \ ,
$$
with a diagonal mass matrix; the case of a general mass matrix can be treated 
similarly. In the main text we quote results assuming all the masses $m_i$ to be 
equal.  We start by replacing
${\bf x}$ by ${\bf x}+\ve$ in the exponent and expanding to second order. 
Requiring the linear terms in $\ve$ to vanish gives the classical equation
of motion as well as the inital condition for ${\bf x}$ (\ref{veq2}). The 
remaining terms give the approximation 
$$
\psi({\bf X},T)
\approx \sum e^{iS[{\bf x}]/\hbar} \psi_0({\bf x}(0) \int D\ve \ e^{Q[\ve]} 
$$
where 
\bean
Q[\ve] &=& \frac{i}{2\hbar} \left( 
\int_0^T 
\sum_{i=1}^d  m_i\dot{\ve}_i(t)^2 - 
\sum_{i=1}^d\sum_{j=1}^d  H(V)_{ij}({\bf x}(t))  \ve_i(t)\ve_j(t) 
dt  \right. \\
&& \left.~~~~~~~ ~~~~~~~~~~~~~~~~~~
   + \sum_{i=1}^d\sum_{j=1}^d H(S^{\rm init})_{ij}({\bf x}(0))  \ve_i(0)\ve_j(0)
\right) \ .
\eean
Here $H(V)$ and $H(S^{\rm init})$ 
denote the matrices of second derivatives of 
$V$ and $S^{\rm init}$ respectively. 
We proceed by 
discretizing the integrals in $Q[\ve]$. Bearing in mind that $\ve(T)=0$ and using 
the trapezium rule we have
\bean 
\int_0^T     
  H(V)_{ij}({\bf x}(t))  \ve_i(t)\ve_j(t) 
dt
&\approx& 
   \frac{T}{2N}    H(V)_{ij}({\bf x}(0))  \ve_i(0)\ve_j(0)  \\
&&  + \frac{T}{N} \sum_{r=1}^{N-1} 
   H(V)_{ij}\left({\bf x}\left(\frac{rT}{N}\right)\right)  
   \ve_i\left(\frac{rT}{N}\right)
   \ve_j\left(\frac{rT}{N}\right) 
\ .  \nonumber
\eean 
Using a forward difference approximation for the derivative of $\ve(t)$ and  
a ''leftbox'' type approximation for the relevant integral gives the approximation 
$$ 
\int_0^T 
\dot{\ve}_i(t)^2 dt 
\approx   \frac{N}{T}\left(
\ve_i(0)^2 + 2 \sum_{r=1}^{N-1} \ve_i\left( \frac{rT}{N} \right)^2 - 2 
\sum_{r=0}^{N-2}  \ve_i\left( \frac{rT}{N} \right)\ve_i\left( \frac{(r+1)T}{N} \right)
\right) \ .
$$
(Although this would appear to be a first order approximation, since both the 
methods for constructing the derivative and computing the integral are first order,
it is actually second order; the first order errors in the methods exactly cancel
each other.) 
Putting this all together gives:
$$
Q[\ve] \approx
\frac{iN}{2\hbar T} \Delta \left( 
\begin{array}{cccccc} 
M+F_0  & -M      & 0      & 0      &   0 & \ldots  \\
-M     & 2M+F_1 & -M      & 0      &   0 & \ldots  \\ 
0     & -M      & 2M+F_2 & -M      &   0 &         \\
0     & 0      & -M      & 2M+F_3 &   -M &         \\
\vdots & \vdots&        &   \ddots     &  \ddots   &      \\
       &       &        &        &     -M  & 2M+F_{N-1} 
\end{array}
\right) \Delta^T \ .
$$
Here $\Delta=\pmatrix{\ve(0) &\ve(T/N)& \ve(2T/N)& \ldots\cr }$. Each entry in the 
matrix in the previous equation is a $d\times d$ block matrix. 
$M$ denotes the diagonal matrix with entries
$m_i$, and the matrices $F_r$ are defined by 
\bean
(F_0)_{ij} &=&  \frac{T}{N}H(S^{\rm init})_{ij}({\bf x}(0)) - 
                \frac{T^2}{2N^2}H(V)_{ij}({\bf x}(0)) \ ,
\\
(F_r)_{ij} &=&  - \frac{T^2}{N^2}H(V)_{ij}\left({\bf x}\left(\frac{rT}{N}\right)\right) 
  \ ,\qquad r=1,\ldots,N-1 \ .
\eean
As a final step in the simplification of $Q[\ve]$, we factor out factors of 
$\sqrt{M}$ on the right and the left from each block in the above matrix. 
In this way we obtain
\be
Q[\ve] \approx
\frac{iN}{2\hbar T} \Delta \sqrt{\cal M} \left( 
\begin{array}{cccccc} 
I+G_0  & -I     & 0      & 0      &   0 & \ldots  \\
-I     & 2I+G_1 & -I      & 0      &   0 & \ldots  \\ 
0     & -I      & 2I+G_2 & -I      &   0 &         \\
0     & 0      & -I     & 2I+G_3 &   -I &         \\
\vdots & \vdots&        &   \ddots     &  \ddots   &      \\
       &       &        &        &     -I  & 2I+G_{N-1} 
\end{array}
\right) \sqrt{\cal M} \Delta^T \ .
\label{bigG}\ee
where the matrix ${\cal M}$ is a diagonal matrix with $N$ copies of $M$ on 
its main diagonal, and the matrices $G_r$ are defined by 
\bean
(G_0)_{ij} &=& \frac1{\sqrt{m_im_j}}\left( \frac{T}{N}H(S^{\rm init})_{ij}({\bf x}(0)) - 
                \frac{T^2}{2N^2}H(V)_{ij}({\bf x}(0)) \right) \ ,
\\
(G_r)_{ij} &=&  - \frac1{\sqrt{m_im_j}}
\frac{T^2}{N^2}H(V)_{ij}\left({\bf x}\left(\frac{rT}{N}\right)\right) 
  \ ,\qquad r=1,\ldots,N-1 \ .
\eean
It remains to compute the determinant of the matrix in (\ref{bigG}). To do this, we 
first apply block Gaussian elimination \cite{Harry} to eliminate the blocks under the 
leading block diagonal, a process that does not affect the determinant. This gives
a matrix of the form 
$$
\left( 
\begin{array}{cccccc} 
P_0  & -I     & 0      & 0      &   0 & \ldots  \\
0    & P_1 & -I      & 0      &   0 & \ldots  \\ 
0     &  0      & P_2 & -I      &   0 &         \\
0     & 0      &  0    & P_3 &   -I &         \\
\vdots & \vdots&        &   \ddots     &  \ddots   &      \\
       &       &        &        &     0  & P_{N-1} 
\end{array}
\right) 
$$
where
\bean
P_0 &=& I+G_0 \ ,\\
P_1 &=& 2I+G_1 - P_0^{-1} \ , \\
P_2 &=& 2I+G_2 - P_1^{-1} \ ,\\
    &\vdots & \\
P_{N-1} &=& 2I+G_{N-1} - P_{N-2}^{-1} \ . 
\eean
We wish to find $\det \left(P_0P_1P_2\ldots P_{N-1}\right)$. Define the matrices 
$R_r$, $r=0,\ldots,N-1$, by $R_r = P_0P_1P_2\ldots P_r$. Then we have 
\bean
R_0 &=& I+G_0 \ , \\
R_1 &=& (I+G_0)(2I+G_1) - I  = I + 2G_0 + G_1 + G_0G_1 
\eean
and for $2\le r\le N-1$: 
$$
R_r = R_{r-2}P_{r-1}P_r 
    = R_{r-2}\left( P_{r-1}(2I+G_r) - I \right) 
    = R_{r-1}(2I+G_r) - R_{r-2} 
$$
Rewriting the last equation, for $2\le r\le N-1$ we have
$$ \frac{ R_r - 2R_{r-1} + R_{r-2} }{(T/N)^2} 
   = R_{r-1}\frac{N^2}{T^2} G_r\ . $$
Taking the limit now as $N\rightarrow\infty$, we see that the sequence of matrices
$R_0,R_1,\ldots,R_{N-1}$ is replaced by a function $R(t)$, defined by 
$$ \ddot{R}(t) = R (t)G(t) \ , \qquad 
{\rm where}~~
G_{ij}(t) =  - \frac1{\sqrt{m_im_j}}H(V)_{ij}({\bf x}(t)) \ ,
$$ 
supplemented with the initial conditions 
$$ R(0)=I \ , \qquad 
\dot{R}_{ij}(0)=\frac1{\sqrt{m_im_j}}H(S^{\rm init})_{ij}({\bf x}(0))\ . $$ 
Finally, we write $U(t)=R(t)^T$. Taking the transpose of all the equations 
above we see that $U(t)$ satisfies 
$$ \ddot{U}(t) = G(t)U(t) \ , \qquad  U(0)=I, ~~
\dot{U}_{ij}(0)=\frac1{\sqrt{m_im_j}}H(S^{\rm init})_{ij}({\bf x}(0))\ . $$ 
The determinant of the matrix in (\ref{bigG}) is simply $\det U(T)$, and (after
checking the case of the free particle to fix the normalization) we deduce 
that 
$$
\psi({\bf X},T)
\approx \sum \frac{ e^{iS[{\bf x}]/\hbar} \psi_0({\bf x}(0) }{\sqrt{\det U(T)}}
\ .$$ 
as required. 


\end{document}